\documentclass[prl,aps,twocolumn,10pt,a4paper]{revtex4}
\usepackage{amsmath}
\usepackage{amsfonts}
\usepackage{amssymb}
\usepackage{epsfig}
\usepackage{wrapfig}

\newcommand{\ud}{\,\mathrm{d}}
\newcommand{\s}{\sigma}
\renewcommand{\P}{\mathbb{P}}
\newcommand{\beq}{\begin{equation}}
\newcommand{\eeq}{\end{equation}}
\newcommand{\ben}{\begin{eqnarray}}
\newcommand{\een}{\end{eqnarray}}

\begin{document}

\title{Synteny in Bacterial Genomes: Inference, Organization and Evolution}

\author{Ivan Junier} 
\affiliation{Centre for Genomic Regulation (CRG), Dr. Aiguader 88, 08003 Barcelona, Spain; Universitat Pompeu Fabra (UPF), 08003 Barcelona, Spain}
\author{Olivier Rivoire}
\affiliation{CNRS/UJF-Grenoble 1, LIPhy UMR 5588, Grenoble, 38402, France}

\begin{abstract}
Genes are not located randomly along genomes. Synteny, the conservation of their relative positions in genomes of different species, reflects fundamental constraints on natural evolution. We present approaches to infer pairs of co-localized genes from multiple genomes, describe their organization, and study their evolutionary history. In bacterial genomes, we thus identify synteny units, or "syntons", which are clusters of proximal genes that encompass and extend operons. The size distribution of these syntons divide them into large syntons, which correspond to fundamental macro-molecular complexes of bacteria, and smaller ones, which display a remarkable exponential distribution of sizes. This distribution is "universal" in two respects: it holds for vastly different genomes, and for functionally distinct genes. Similar statistical laws have been reported previously in studies of bacterial genomes, and generally attributed to purifying selection or neutral processes. Here, we perform a new analysis based on the concept of parsimony, and find that the prevailing evolutionary mechanism behind the formation of small syntons is a selective process of gene aggregation. Altogether, our results imply a common evolutionary process that selectively shapes the organization and diversity of bacterial genomes.
\end{abstract}

\maketitle

The position of genes along genomes affect their function and evolution. In bacteria, functional constraints are thus responsible for the concentration of highly expressed genes near the origin of replication, and the clustering of co-functional genes into operons of co-regulated genes. Similarly, evolutionary constraints on gene order are evidenced in bacteria by the highly variable rates of recombination over different chromosomal regions, and by the propensity of co-localized genes to be co-displaced through horizontal transfer~\cite{Lawrence2003}. Thus, while genes may be lost, gained, duplicated and rearranged during evolution, the comparison of evolutionary related species shows a remarkable stability of genomic organization~\cite{Koonin:2008id,Rocha:2008jr}.

This conservation of genomic organization, often referred to as "synteny", raises three questions: (i)~How to infer, from a comparison of multiple genomes, the pairs of genes with significant conservation of proximity? (ii)~How to describe, beyond pairwise relationships, the organization of conserved properties of co-localization? (iii)~How to explain, from an evolutionary perspective, the origin of this organization?

Following upon several previous studies~\cite{LatheIII:2000um,Tamames:2001wk,Rogozin:2002ti,Snel:2002bo,Rocha:2005bm,Wright:2007dv,Fang:2008iv}, we tackle these questions through a multi-genome comparative analysis, using as input the over one thousand complete bacterial genomic sequences that are now available. Our approach to answering (i) yields, at each given phylogenetic level, a list of pairs of genes that remain close-by across multiple genomes. In the spirit of \cite{Rogozin:2002ti}, we address (ii) by projecting these pairs on individual genomes to define synteny units, or syntons, as clusters of mutually proximal genes. We find that these syntons correspond to functional features of bacterial genomes encompassing operons, and that the distribution of their size partition them in two classes: large syntons, associated with fundamental functions of bacterial cells, and smaller ones, which in many cases do not correspond to previously defined genomic units. These smaller syntons, however, follow a remarkable statistical property, not previously reported, with their sizes being exponentially distributed.

Finally, in response to (iii), we note that two types of evolutionary processes can account for the conservation of genomic organizations over different clades : (1) an incomplete dislocation of gene order from a common ancestral genome \cite{Nadeau:1984tm, Rocha:2004bx}, or (2) a selective aggregation of genes to form new clusters. These two processes are not exclusive, and their balance may for instance explain the conservation of the larger syntons~\cite{Fang:2008iv}. Both processes are also compatible with the exponential size distribution for the smaller syntons; however, we provide evidence that  the accumulation of genes into expanding syntons is the dominating process, a result in contrast with previous models that considered disintegration as the driving force of synteny evolution~\cite{Rocha:2005bm}.

\subsection{Inference of synteny}

Inferring synteny from multiple genomes presents several difficulties: (a) the classification of genes into orthology classes, which is a notoriously difficult problem~\cite{Gabaldon:2013fi}; (b) an highly non-uniform sampling of genomes, both because natural genomes are phylogenetically related and because sequencing efforts have not been distributed evenly across strains and species; (c) the definition of a non-ambiguous criterion for assessing significant conservation of proximity between genes. Our approach specifically addresses each of these difficulties: (a) given an initial partition of genes into orthology classes, we use the inferred properties of synteny as a guide to iteratively refine it, and to eventually test the consistency of our results; (b) one parameter in the analysis allows for the investigation of syntenic properties at different phylogenetic depths, and can thus mitigate biases from closely related genomes; (c) our thresholds of statistical significance are defined for any given rate of false positive discovery of conserved proximities.

The input data is a set of $M$ genomes whose genes are partitioned into $N$ orthologous classes. Specifically, we consider here a set of $M=1108$ bacterial genomes annotated in terms of $N=4467$ clusters of orthologous genes (COGs; see Methods). COGs are defined from gene sequences only, with no reference to gene positions, based on the principle that any group of at least three genes from distant genomes that are more similar to each other than to any other genes from the same genome should belong to the same COG~\cite{Tatusov:2000tu}. As a result, a genome may contain one, several or no gene associated to any given COG.

Available complete genomic sequences of bacteria are phylogenetically related and unevenly sampled. Treating them equivalently therefore amounts to biasing the statistics towards the most represented species and clades. To correct for this bias, we  underweight each genome in proportion to the number of other genomes to which it is similar, thus naturally defining an effective number of genomes $M'\leq M$~\cite{Morcos:2011jg} (Methods). This approach introduces one parameter, the evolutionary distance $\delta$ below which two genomes are considered to be similar (Fig.~\ref{fig:Fig1}A). As we are interested in synteny, we take here for $\delta$ a measure of divergence of gene contexts (Figs.~S1-S2). Varying $\delta$ allows us to perform the same analysis at different phylogenetic levels, thus providing an information that is either generic to many bacteria (large $\delta$), or specific to a small subset of them (small $\delta$).

For each pair of COGs $ij$, we define its relative distance in a genome as the minimal distance, in base pairs, between its respective genes (this distance is formally $\infty$ if one of the COGs is not represented in the genome). The distribution of this distance across all genomes is computed by taking into account the $\delta$-dependent genome weights. We then assign a $p$-value $\hat \pi_{ij}$ to the pair $ij$ by comparing this distribution with that from a null model where genes are distributed independently and uniformly across $M'$ genomes (Fig.~\ref{fig:Fig1}, Methods and Fig.~S4).

Given the large number of pairs $ij$ under study ($\sim 10^7$), some of the $p$-values $\hat \pi_{ij}$ are borne out to be very small, even under the null model. One more step is therefore required to set a threshold of significance for these $p$-values. This is achieved by comparing the empirical distribution $f(\pi)$ of $\hat\pi_{ij}$ with its distribution under the null model, $f_0(\pi)$. The fraction of false positives when calling significant the pairs $ij$ with $\hat\pi_{ij}<\pi^*$ can be estimated as $q=\int_0^{\pi^*}f_0(\pi)\ud\pi/\int_0^{\pi^*}f(\pi)\ud\pi$  (Methods). A given false discovery rate $q$ (the fraction of false positives given $\pi^*$), here taken at 5\%, thus selects a threshold of significance $\pi^*$~\cite{Benjamini:1995ws}. This procedure corresponds to applying to synteny properties the approach of~\cite{Storey:2003cj}, with a simpler but more stringent criterion justified by the small fraction of true positives.

\subsection{Organization of synteny}

The structure of the relationships of conserved proximity can first be analyzed on individual genomes, in line with the work of Rogozin {\it et al}~\cite{Rogozin:2002ti}. To this end, we assess whether each pair $ij$ of COGs found to be significantly proximal, is indeed close-by in the particular genome $g$, using $\hat x_{ij}$ as a characteristic distance (Methods and Fig.~S5). This defines a genome-based network of proximity, where the nodes are genes and the links relations of conserved proximity in $g$. To identify synteny units from this network, we rely on a property of transitivity: for $ijk$ to be considered as a unit, all three pairs $ij$, $ik$ and $jk$ must be linked. This corresponds to relevant sub-networks being fully inter-connected, so-called "cliques" in graph theory (Fig.~\ref{fig:Fig2}A). We thus define the "syntons" of a genome as the maximal cliques of its network of proximity, i.e., the cliques that are not strictly contained in any other. The syntons of a genome hence consist of maximal sets of genes that are proximal in the genome as well as in a significant number of other genomes.

Syntons are defined by comparing the positions of genes between species, without taking into account promoters or terminators. Yet, while not necessarily consisting of contiguous genes, syntons are related to operons. As shown in Fig.~\ref{fig:Fig2} for the {\it E. coli} genome~\cite{GamaCastro:2011if}, the partitions into syntons and operons share many of their boundaries; besides, operons are rarely found in two different syntons. These results indicate that syntons comprise operons but extend beyond them, just as uber-operons~\cite{LatheIII:2000um}, superoperons~\cite{Rogozin:2002ti}, persistent genes \cite{Fang:2008iv}, clusters of pathway-related operons \cite{Yin:2010jq} and statistically correlated genes \cite{Junier:2012dc}.

The size distribution of syntons is particularly informative: as shown in Fig.~\ref{fig:Fig3}, it displays a critical size that defines two distinct types of syntons. Syntons above this size (whose exact value depends on the phylogenetic depth $\delta$ of the analysis) are found to contain genes encoding the building blocks of the fundamental molecular complexes of bacterial cells, including the translation/transcription machinery, the ATP-producing respiratory complex, the cell division complex, the cell envelope biogenesis and the flagellum machinery (Table~S1). We refer to these syntons as syntons of type  B (with 'B' for 'basic building blocks').

After removing these syntons, the size distribution of the remaining syntons follows an exponential law, $\rho(\s) \sim e^{-\alpha \s}$, with an exponent $\alpha$ that varies slightly with the phylogenetic depth $\delta$ at which the analysis is performed; the larger $\delta$, the larger $\alpha$ is (Fig.~\ref{fig:Fig3}B), reflecting the fact that smaller and smaller contexts are recognized as conserved when considering wider and wider phylogenetic ranges ($\alpha$ is inversely related to the mean size of syntons). At a given $\delta$, however, the  exponent $\alpha$ is nearly the same for all genomes, irrespectively of phylogenetic distances or genome lengths, and despite the fact that synton compositions may have little overlap (Fig.~\ref{fig:Fig4}). Randomizing the COGs confirms that the exponential law is a {\it bona fide} property of the data, and not a necessary consequence of the methods (Fig.~S6). Hereafter, we refer to the small, exponentially distributed syntons, as syntons of type A (with 'A' for 'aggregating', see below).

Synteny units may also be defined for the "pan-genome" that comprises all bacterial genomes, by considering a pan-network of proximity where the nodes are COGs and the links connect pairs of COGs $ij$ for which $\hat\pi_{ij}\leq\pi^*$ (as in Fig.~\ref{fig:Fig1}B). From this pan-network, partially represented in Fig.~\ref{fig:Fig5}A, we may invoke again transitivity to define as synteny units its maximal cliques. Since forming a clique in this pan-network is, however, only a necessary condition for a set of genes to be proximal in a particular genome, we call them "pan-syntons" only when they are a synton in at least one specific genome.

The number of pan-syntons thus defined is larger than the number of syntons in any genome, which allows for broader statistical tests. For instance, we verify in Fig.~\ref{fig:Fig5}B that the clustering of COGs into pan-syntons is consistent with their annotation into 24 functional categories~\cite{Tatusov:2000tu}. From their size distribution, the same distinction can again be made between type B pan-syntons, associated with macro-complexes, and smaller type A pan-syntons, which sizes are again exponentially distributed. This distribution is preserved when the threshold of statistical significance is varied (Fig.~S7). It also does not appear to be associated with any particular subclass of genes (Fig.~S8). For the same reason that the exponent $\alpha$ decreases with decreasing $\delta$ (Fig.~\ref{fig:Fig3}B), $\alpha$ for the pan-genome (Fig.~\ref{fig:Fig5}C) is smaller than the typical $\alpha$ for a genome studied at same phylogenetic depth $\delta$: the syntons in individual genomes are indeed included in pan-syntons. The fact that all these networks, either global or genome-based, obey, up to a scaling factor, to the same statistical law (Fig.~\ref{fig:Fig5}D), may reflect the "fractal structure of the gene universe"~\cite{KooninBook}.

Finally, we note that $\alpha$ could depend on the number of genomes considered for the analysis, with a larger number of genomes possibly yielding a smaller exponent. Structural constraints on synteny properties \cite{Junier:2012dc}, together with the limited number of COGs shared by many bacteria \cite{Koonin:2008id}, should, however, limit the maximal size of type A syntons, and hence put a bound on $\alpha$. Repeating our analysis for smaller sets of bacterial genomes yields results consistent with these arguments (Fig.~S9).

\subsection{Evolution of synteny}

The organization of genomes changes during evolution as a result of gene losses, acquisitions by duplication or horizontal transfers, and transpositions. The finding of an exponential law for the sizes of small syntons strongly constrains the way in which these different factors must interplay. Yet, qualitatively very different scenarios are consistent with this law: (i)~as for the Boltzmann law in statistical mechanics, it may reflect an equilibrium with conserved mean size of the syntons; (ii)~it may result from the disintegration of an ancestral genome; or (iii)~from a process of aggregation of genes.

Scenario (ii) underlies previously proposed models for rearrangements of genes in eukaryotes~\cite{Nadeau:1984tm} and bacteria~\cite{Rocha:2004bx}. In the simplest such model, two consecutive genes are disrupted at a constant rate $r_d$, leading to a probability $e^{-r_dT\s}$ for $\s$ initially consecutive genes to preserve their integrity across a period of time $T$.

The scenario (iii) of aggregation has, to our knowledge, not been previously considered. In one of its simplest instantiations, rearrangements lead isolated genes, formally forming syntons of unit size, to join the neighborhood of an existing synton of size $k$, and, when this rearrangement confers a selective advantage, it is fixed and generate a new synton of size $k+1$. If $N_k$ denotes the number of syntons of size $k$ in a particular genome, this simple model is described by 
\begin{eqnarray}
\partial_t N_1 &=&\Phi-r_a (1+\rho_1) N_1,\\
\partial_t N_k &=&r_a \rho_1 N_{k-1}-r_a \rho_1 N_k\quad (k\geq 2),
\end{eqnarray}
where $r_a$ is an aggregation rate, accounting for both rearrangement and fixation, $\rho_1=N_1/\sum_{\ell\geq 1} N_\ell$ the density of isolated genes, and $\Phi$ a flux of gene innovation, standing for gene duplications, horizontal transfers or {\it de novo} gene births. Provided that $r_a$ and $\Phi$ vary slowly enough compared to the composition of syntons, this generic model leads to an exponential distribution of cluster sizes.

The evolution of the densities $\rho_k=N_k/\sum_{\ell\geq 1} N_\ell$ of $k$-clusters is indeed given by
\begin{eqnarray}
\partial_t \rho_1 &=&\phi-(r_a+\phi) \rho_1,\label{eq:rho1}\\
\partial_t \rho_k &=&r_a \rho_1 \rho_{k-1}-\phi \rho_k\quad (k\geq 2),\label{eq:rho2}
\end{eqnarray}
where $\phi\equiv\Phi/\sum_{\ell\geq 1} N_\ell$. Assuming that the synteny rates $r_a(t)$ and the flux $\phi(t)$ have their own dynamics on a time scale longer than the time scale of the dynamics described by Eqs.~\eqref{eq:rho1}-\eqref{eq:rho2}, the variables $\rho_k(t)$ reach stationary values before $\phi$ and $r_a$ undergo any noticeable changes. This adiabatic approximation reduces the number of parameters from 2 ($r_a$ and $\phi$) to 1 ($r\equiv\phi/r_a$), and gives as a stationary solution an exponential distribution of synton sizes,
\beq\label{eq:rhok}
\rho_k=\frac{r}{(1+r)^k},\qquad{\rm with}\quad r=\frac{\phi}{r_a}.
\eeq

Without seeking to infer a precise model for synton formation, we may identify the nature of the prevalent scenario, disintegration (ii) or aggregation (iii), by comparing genomic contexts and invoking a principle of parsimony. This principle has been invoked in several studies of genome evolution~\cite{Snel:2002bo,Mirkin:2003ka,Kummerfeld:2005ga}, but, at variance with these previous works, we do not rely here on the reconstruction of a global species tree. Instead, we shall consider triplets of equidistant genomes.

Starting with a pair of genomes $(g_1,g_2)$ sharing a common gene $i$, if $g_1$ has gene $j$ in the context of $i$ but $g_2$ not, two parsimonious explanations are conceivable: $j$ was next to $i$ in the last common ancestor and $g_2$ underwent a disaggregation, or it was not, and $g_1$ underwent an aggregation (Fig.~\ref{fig:Fig6}A). We may attempt to estimate the corresponding probabilities $p_A$ of aggregation ($j$ joining the context of $i$) and $p_D$ of disaggregation ($j$ leaving the context of $i$) over all pairs of genomes separated by a given phylogenetic distance $\delta_s$ (here measured from sequence similarity; Methods). These probabilities are indeed related to the fractions $f_0$, $f_1$ and $f_2$ of those pairs of genomes that both contain a COG $i$, and for which a neighbor $j$ of $i$ in the pan-network of proximity  (Fig.~\ref{fig:Fig5}A) is, respectively, not in the context of $i$ in any of the two genomes, present in one of them, or present in both. The relation, however, also involves the probability $q$ for the ancestor to have $j$ in the context of $i$ and, therefore, cannot define uniquely $p_A$ and $p_D$ (Fig.~\ref{fig:Fig6}A).

This indeterminacy is lifted when considering triplets of equidistant genomes (Fig.~\ref{fig:Fig6}B; practically, two distances $\delta_s$ are considered equal when they differ by $\Delta\delta_s\leq 0.012$). If, besides, conditioning to the presence of a $k$-clique instead of conditioning to the presence of a single COG $i$ (corresponding to $k=1$), we can estimate the rates of aggregation and disaggregation of a gene as a function of the size $k$ of the group of co-localized genes that it is joining or leaving. The results of this analysis, shown in Fig.~\ref{fig:Fig4}C, indicate that $p_A$ is larger than $p_D$ for $k > 2$, all the more that $\delta_s$ is large. The same analysis can be repeated without the condition that $i$ and $j$ must be significantly conserved; in this case, compared to $p_D$, $p_A$ is negligible for all values of $\delta_s$ (Fig.~S11), in agreement with the observation that, except for a small subset of them, genome neighborhoods are poorly conserved among distantly related bacteria \cite{Tamames:2001wk}. For syntons, however, we conclude that aggregations dominate over disaggregations, thus ruling out an equilibrium-like distribution {\it \`a la} Boltzmann as an explanation for the exponential distribution.

\subsection{Discussion}

We presented statistical approaches to infer conserved proximal relationships between genes, identify the relevant units that they are forming, and deduce the nature of evolutionary process behind their formation. Our approaches can be extended to other aspects of genomic organization, and, in the case of the analysis over triplets of equidistant genomes, to other evolutionary processes, such as horizontal transfers or gene duplications. Accounting for these processes will refine the model of synton formation beyond the dichotomy aggregation/disaggregation. At this stage, we note that while our results imply the existence of a selective pressure for aggregating genes to stick together, they do not reveal its nature. We can only point out that the probability of gene aggregation, $p_A$, appears to be nearly independent of the size of the synteny unit already present in the genomes ($k$ in Fig.~\ref{fig:Fig6}C). 

Besides the inference of evolutionary processes, our results have implications for the functional interpretation of genomic sequences. Synteny properties are indeed commonly used to discriminate paralogs~\cite{Mering:2003hp}, based on the premise that similarity of context correlates with similarity of function. Genomic contexts can thus refine the partition into COGs, which is based on sequences only, by defining smaller clusters of genes with analogous contexts~\cite{Yelton:2011is}. Following this line, we derived from our results a set of contextually refined COGs, which we call cCOGs (Methods). Repeating our analysis with the cCOGs as fundamental units consistently reinforces the results found from the COGs. All the cliques of the corresponding pan-network of proximal relationships now consist of genes that are co-localized in at least one genome: pan-syntons can therefore be defined as maximal cliques of the pan-network of proximity with no further condition. From the size distribution of syntons, the same distinction can be made between syntons of types A and B, and the exponential distribution for the syntons of type A is only more significant (Fig.~S12).

\subsection{Conclusion}

In conclusion, we presented a statistical study of multiple bacterial genomes that leads to the identification of novel units of synteny, called syntons. Identifying the relevant units of coevolution between genes is an essential step towards the rationale design of {\it de novo} functional genomes \cite{Kepes:2012kq}. In addition, the exponential distribution of synton sizes may be added to the list of statistical "laws of genome evolution"~\cite{Luscombe:2002wc,vanNimwegen:2003if,Wolf:2009jy,Molina:2009go,Grilli:2012ka,Pang:2013ea}. Previous examples could be explained by neutral processes and/or purifying selection~\cite{Koonin:2011in}. The exponential distribution of synton sizes is remarkable for being driven by a process of positive selection, an on-going accretion of genes.

Genomic features other than synteny, such as for instance the co-occurrence of orthologous genes, may be analyzed along the same lines~\cite{JRprep}. As these features may result from an evolutionary dynamics of a different nature than the aggregative process leading to syntons, the relevant subunits of the network of conserved properties may not be its maximal cliques and/or their sizes may not be exponentially distributed. Identifying those relevant conserved units, studying their properties, and inferring the evolutionary mechanisms behind their formation are avenues for future studies of bacterial evolution.

\section{Materials}

\indent {\bf Data set} -- Sequenced bacterial genomes and COG annotations were downloaded from NCBI, yielding an initial data set of $M_0=1432$ genomes and $N=4467$ COGs. We removed genomes with size below 500 kb or with less than 60 \% of genes annotated by COGs to obtain the $M=1108$ genomes used in our analysis.\\

\begin{figure*}[c]
\begin{center}
\includegraphics[width=.5\linewidth]{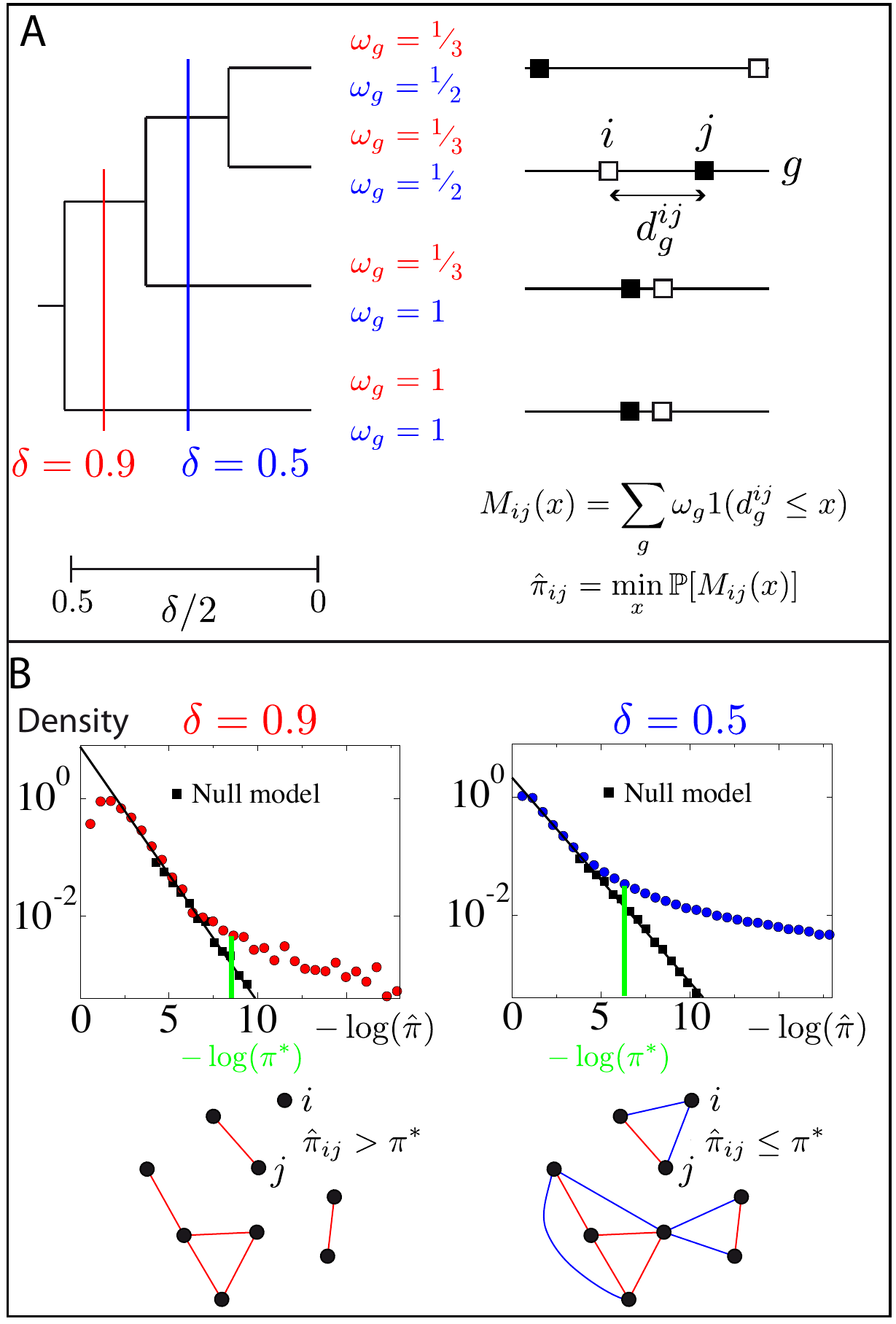}
\caption{{\bf ({A})} A weight $w_g$ is defined for each genome $g$ that is inversely proportional to the number of other genomes at distance less than $\delta$ from it. This distance is defined from the divergence of gene contexts between genomes, and $M'=\sum_g w_g$ gives the effective number of genomes at this phylogenetic level (in this illustration, 2 at $\delta = 0.9$ and 3 at $\delta=0.5$). For a given $\delta$, we compute for every pair of COGs $ij$ the effective number $M_{ij}(x)$ of genomes for which the genomic distance $d^{ij}_g$ separating their genes is less than $x$. This number is converted into a $p$-value, $\P[M_{ij}(x)]$, by considering a null model where gene positions are independently and uniformly distributed in $M'$ genomes (Methods). Finally, the significance of the co-localization between $i$ and $j$ is defined by the minimum $\hat \pi_{ij}$ of these $p$-values over the distances $x$. {\bf (B)} The tail of the distribution of $-\log(\hat \pi_{ij})$ is exponential under the null model (black dots). The empirical distributions (red and blue) clearly deviate from it for small values of $\hat\pi_{ij}$ (large values of $-\log\hat\pi_{ij}$). We select a threshold of significance $\pi^*$ (green) such that the false discovery rate, estimated from the areas below the two curves (Methods), is less than 5\%. For smaller $\delta$, more relationships of conserved proximity are thus detected, as depicted here by the (pan-)networks of proximity, where nodes represent COGs and links relationships of proximity  (the red links at $\delta = 0.5$ are those already identified at $\delta = 0.9$).
\label{fig:Fig1}}
\end{center}
\end{figure*}

\begin{figure*}[t]
\begin{center}
\includegraphics[width=.7\linewidth]{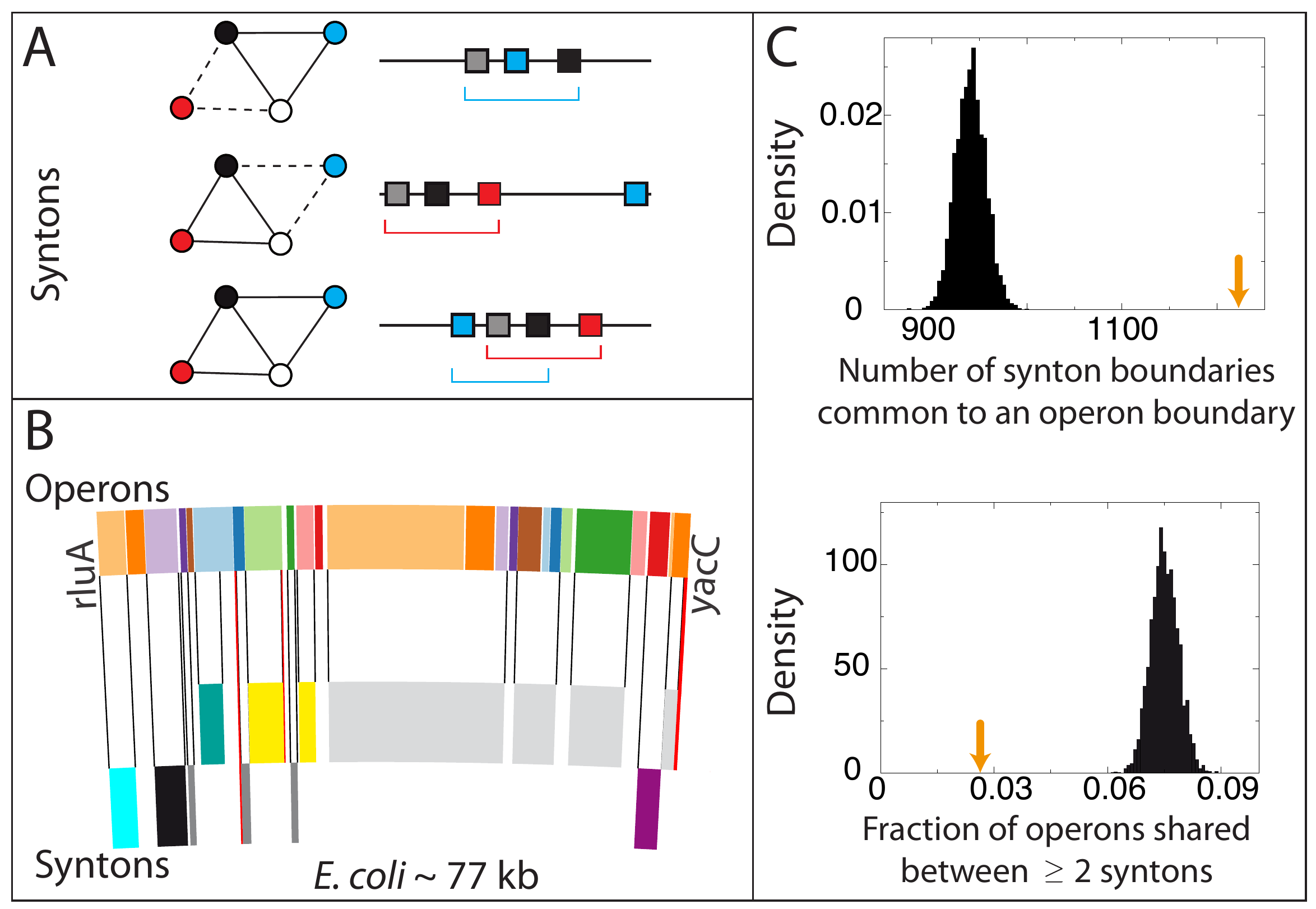}
\caption{{\bf (A)} A synton is a maximal set of genes that are proximal both globally across genomes (dotted lines) and in a particular genome (full lines). The case of three different genomes is depicted here, which each have different syntons. In the last case, the genome has two overlapping syntons because the blue and red genes have no conserved proximity (the four genes do not form a clique in the pan-network). {\bf (B)} Comparison between synton and operon organizations in {\it Escherichia coli}, here for $\delta = 0.53$ ($M'=469$). Only a small fraction (77kb) of the {\it E. coli} genome is represented, with the operons in the upper band and the syntons in the lower bands (two bands are used for clarity; overlapping syntons have the same colors). By definition, operons are made of consecutive genes; in contrast, syntons may intermingle (as it is the case here for the yellow and dark grey syntons). The extreme boundaries of the operons are, however, often common to the synton splittings and synton separations (indicated by thin black lines). Many apparent mismatches come from an absence of COG annotation (red lines). {\bf (C)} Statistical significance of these observations (see Fig. S5 for other values of $\delta$). The first histogram shows the expected number of common boundaries between operons and syntons after a random rotation of the operon organization (hence keeping all other features intact), with the orange arrow indicating the observed value. The second histogram shows the fraction of operons containing genes in at least two different syntons, with again the orange arrow pointing to the observed value. These graphs indicate that syntons are genomic features encompassing operons.
\label{fig:Fig2}}
\end{center}
\end{figure*}

\begin{figure*}[t]
\begin{center}
\includegraphics[width=.7\linewidth]{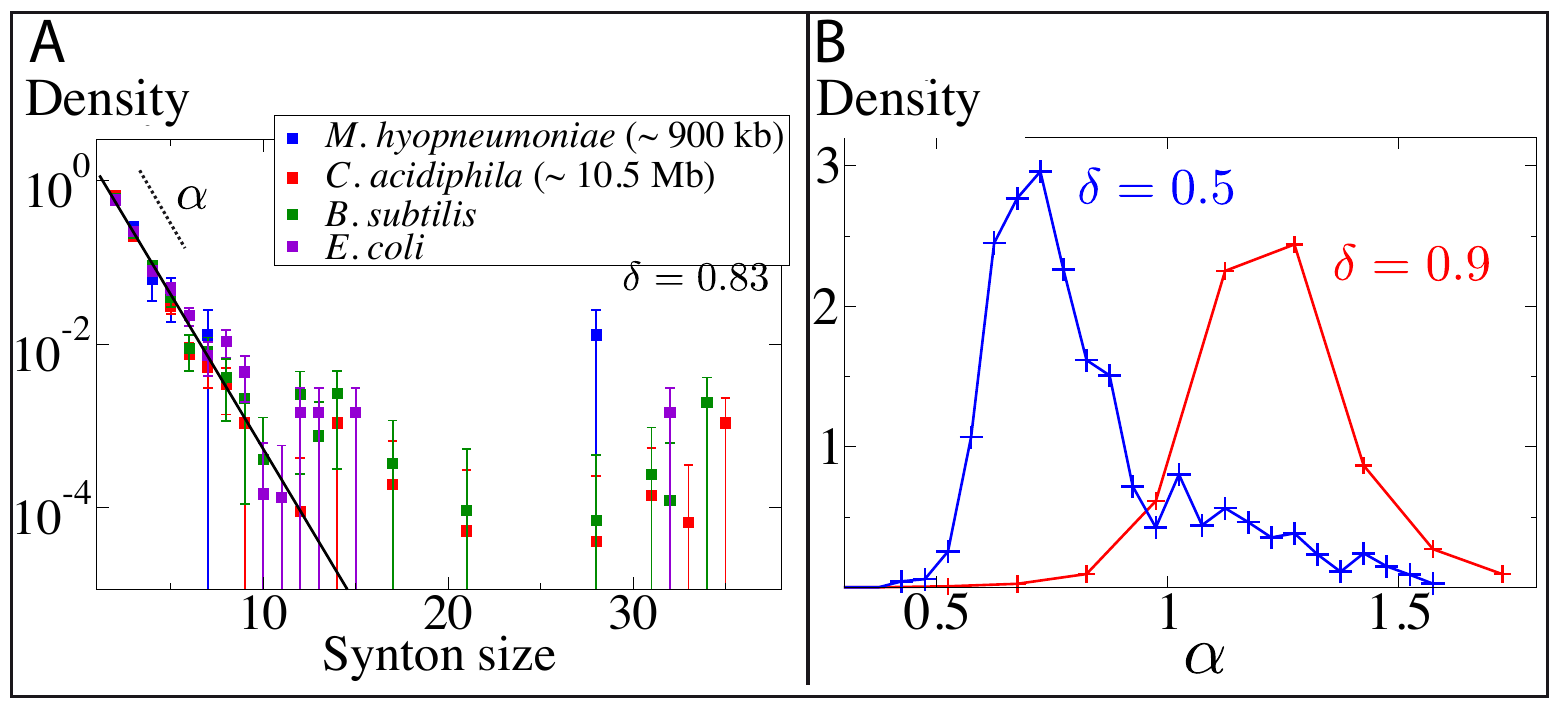}
\caption{{\bf (A)} For a given $\delta$ (here $=0.83$), the size distribution of syntons is similar for genomes that are phylogenetically far apart, or of vastly different sizes (e.g. \emph{Mycoplasma hyopneumoniae}, 900 kb long genome, green points, versus \emph{Catenulispora acidiphila}, 10.5 Mb long, orange points). These distributions can be divided into two parts: above a critical size (around 10 here), the (type B) syntons contain genes encoding the building blocks of the fundamental molecular complexes of bacterial cells. Below this critical size, the (type A) syntons have their sizes distributed according to an exponential law. {\bf (B)} The exponent $\alpha$ of the exponential is nearly identical for all genomes, but varies with the phylogenetic depth $\delta$ at which conservation is estimated.
\label{fig:Fig3}}
\end{center}
\end{figure*}

\begin{figure*}[t]
\vspace*{.05in}
\centerline{\includegraphics[width=.6\linewidth]{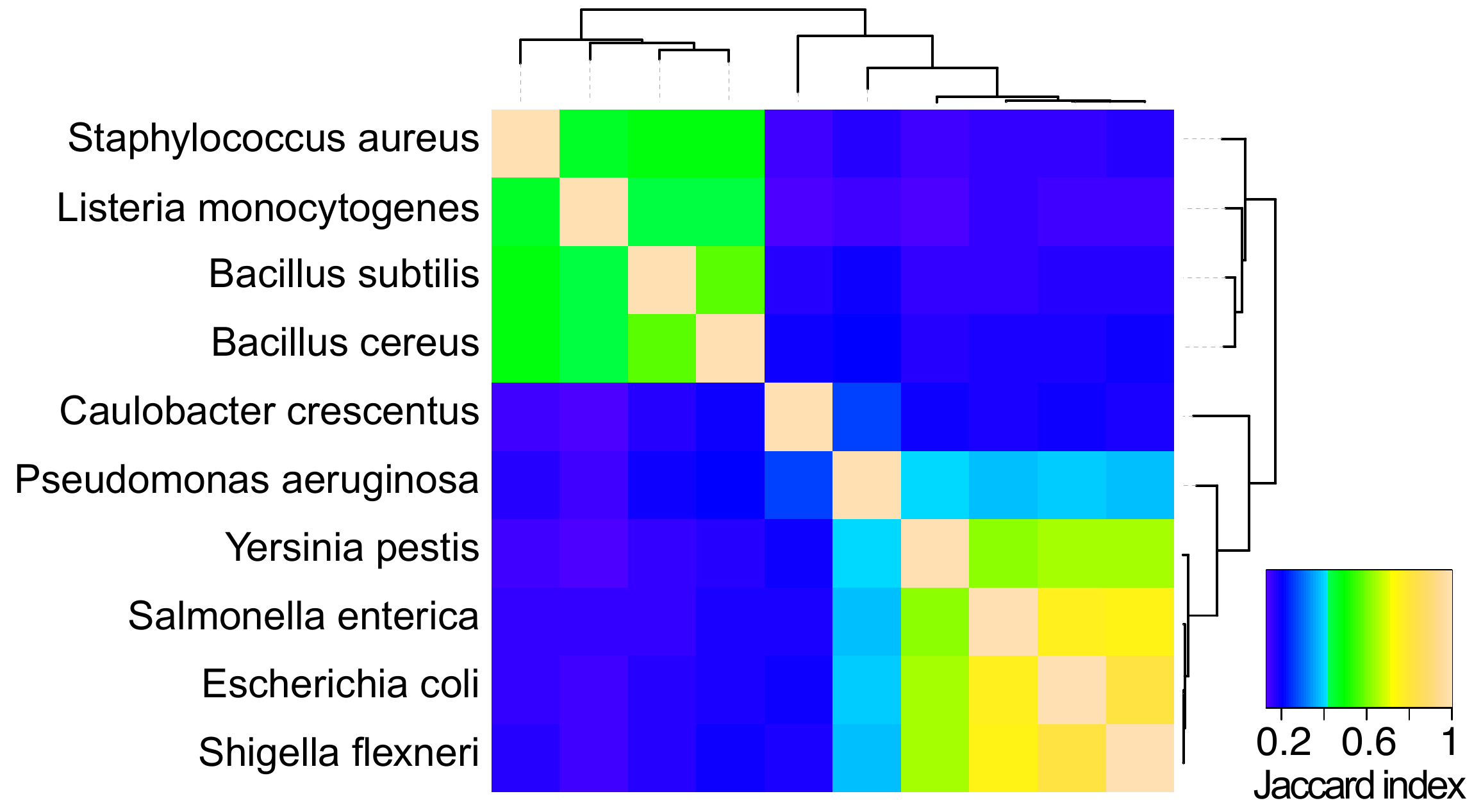}}
\caption{Comparison of synton content between bacteria. The similarity in synton content between two species is defined as their number of common syntons divided by their total number of different syntons (Jaccard index). These similarities are shown here for a sample of 10 strains, showing a clear correlation with the phylogenetic relationships between species, as defined from sequence similarity and indicated by the trees.
\label{fig:Fig4}}
\end{figure*}

\begin{figure*}[t]
\begin{center}
\includegraphics[width=.75\linewidth]{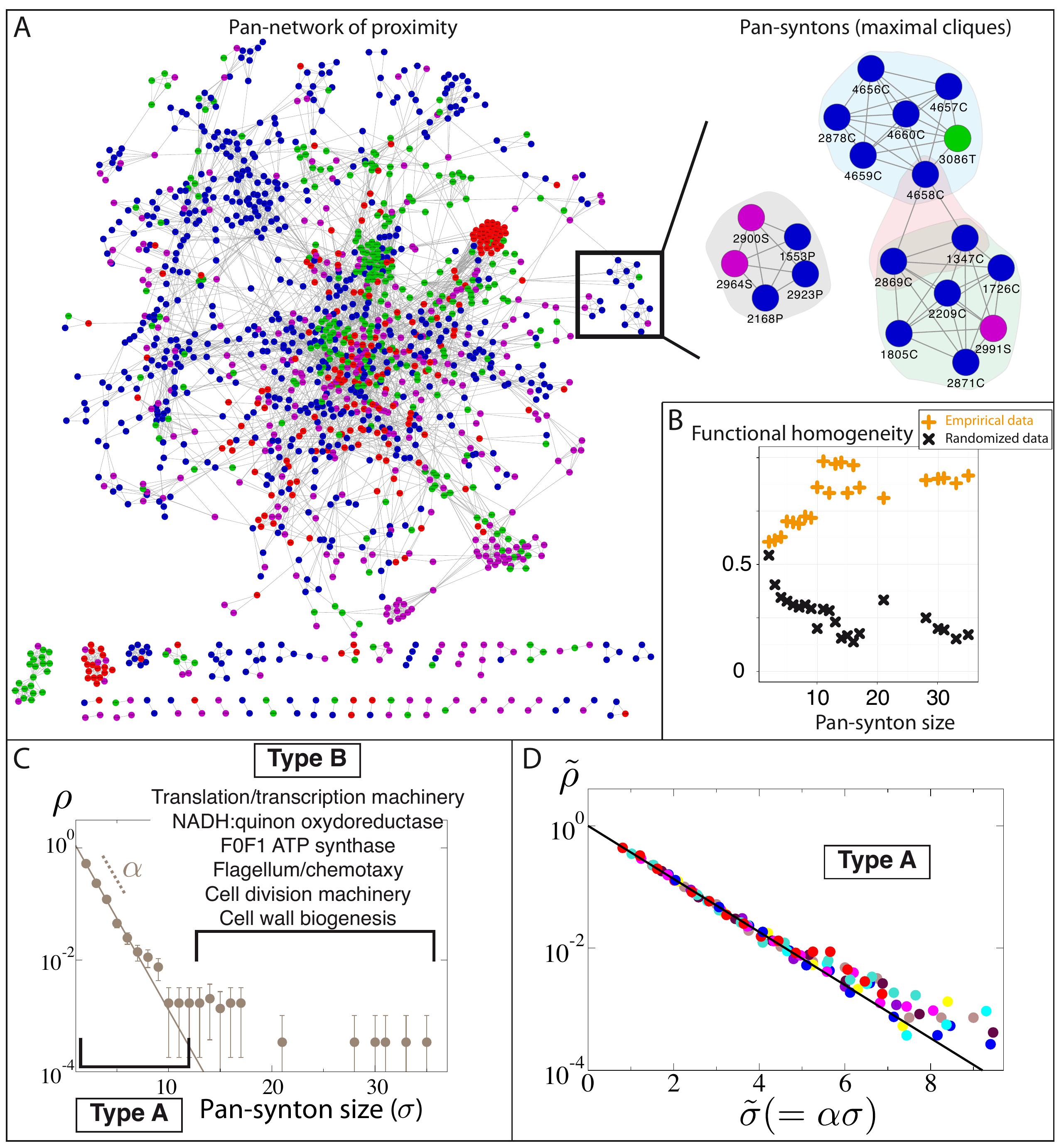}
\caption{{\bf (A)} Pan-network of pairs of COGs with significantly conserved proximity. Here, only a subset of this pan-network, comprising 1455 COGs, is shown ($\delta = 0.83$). Colors indicate four functional classes of COGs: red for information storage and processing, green for cellular processes and signaling, blue for metabolism and purple for poorly characterized. On the right, 19 COGs are highlighted, which are divided into 4 pan-syntons, with one pan-synton  containing 3 COGs (transparent red) that belongs to two other disjoint pan-syntons (transparent green and blue). Maximal cliques (pan-syntons) were computed using CFinder~\cite{Palla:2005cj}. {\bf (B)} Maximal fraction of common function shared by the COGs forming a pan-synton, among the 24 functions defined in the NCBI database~\cite{Tatusov:2000tu}. The figure reports the average of these values over pan-syntons of given sizes, with black crosses indicating the values obtained after reshuffling the COG labels in the pan-network. {\bf (C)} The size distribution of pan-syntons quickly decreases up to a critical size that depends on $\delta$ (around 10 for $\delta = 0.83$). Type B pan-syntons, defined as pan-syntons above this size, are found to contain genes encoding the building blocks of the fundamental molecular complexes of bacterial cells. Type A pan-syntons, below the critical size, have their sizes distributed according to an exponential law, with an exponent $\alpha$ that depends weakly on the phylogenetic depth ($\alpha \simeq 1$ at $\delta= 0.9$, $\alpha \simeq 0.4$ at $\delta = 0.5$). The error bars are standard errors of the mean. Rescaling the size $\sigma$ of pan-syntons for different $\delta$ by $\tilde \sigma \equiv \alpha \sigma$ leads to distributions $\tilde \rho$ that collapse into a single exponential; deviations at the largest values are within error bars (not indicated for clarity).
\label{fig:Fig5}}
\end{center}
\end{figure*}

\begin{figure*}[t]
\begin{center}
\includegraphics[width=.6\linewidth]{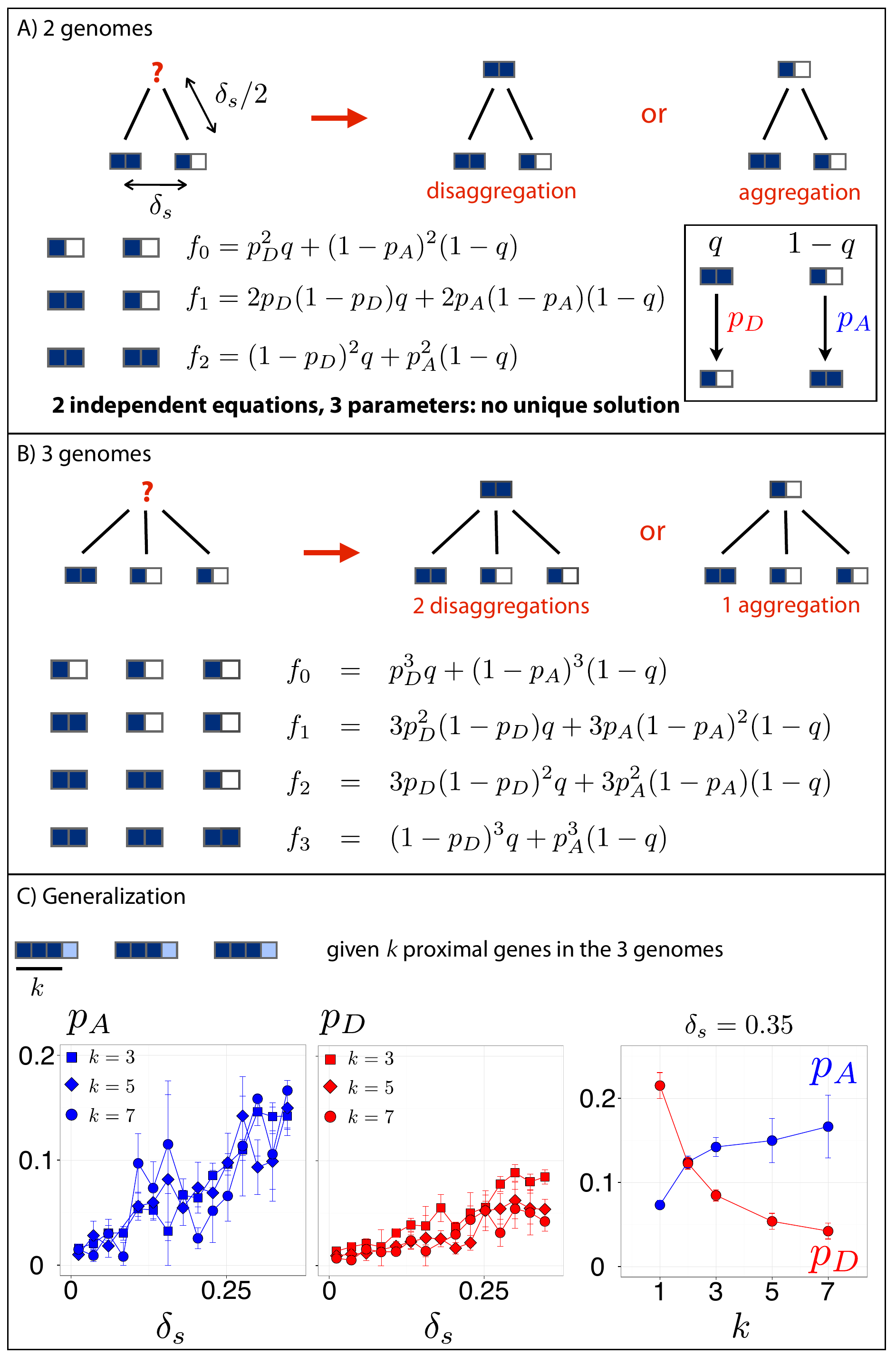}
\caption{
Two non-exclusive mechanisms can explain the exponential size distribution of syntons: aggregation or disaggregation. {\bf (A)} Given a COG $i$, we can compare its context in all pairs of genomes that contain it. For each COG $j$ significantly co-localized with $i$, we compute the fraction $f_n$ of pairs of genomes at distance $\delta_s$ where $n$ out of the two genomes have $j$ in the context of $i$ ($n=0,1,2$). In a simple model where events of aggregation or disaggregation occur with probabilities $p_A$ and $p_D$ since the divergence between the two genomes, these frequencies are related to the probability $q$ that $j$ belongs to the context of $i$ in the last common ancestor of the pair. As $f_0+f_1+f_2=1$, this gives only two independent equations for three unknown parameters. {\bf (B)} The same approach, but based on triplets of equidistant genomes, now provides three independent equations for the same three unknown parameters. {\bf ({C})} A further generalization consists in fixing a clique of $k$ proximal genes instead of a single gene $i$. We thus report $p_A$ and $p_D$ for $k=3,5,7$ as a function of the distance $\delta_s$ between the genomes, and for $\delta_s = 0.35$ as a function of $k$ (the error bars are standard errors). The last graph shows that $p_A>p_D$ for $k>2$. Fig.~S10 extends these results to all values of $k \in [1,8]$ and $\delta_s \in [0,0.35]$.
\label{fig:Fig6}}
\end{center}
\end{figure*}

\indent {\bf Inter-genome distances and genome weights} -- A measure of distance $D_{gh}$ between pairs of genomes is defined from the divergence of contexts of 10 genes known to be vertically inherited in bacterial genomes~\cite{Zeigler:2003hg} (see Suppl.~Info.~for details). The number $M_{ij}(x)$ of genomes in which genes $i$ and $j$ are at distance $d_{ij}\leq x$ is computed as $M_{ij}(x)=\sum_g\omega_g1(d_{ij}\leq x)$, with genome weights defined by $\omega_g= 1/|\{h:D_{gh}<\delta\}|$~\cite{Morcos:2011jg}; this weighting procedure defines an effective number of genomes as $M'=\sum_g  \omega_{g}$ (Fig.~\ref{fig:Fig1}A).
An alternative measure $\delta_s$ of inter-genome distance is defined from sequence similarity by considering the same 10 genes, and computing the fraction of amino acids that they have in common after aligning them. $\delta_s$ and $\delta$ are correlated (Fig. S2), but because sequences are more conserved than contexts, $\delta_s$ is more appropriate for large divergences.\\

\indent {\bf Significance of proximity} -- 
Assuming a uniform distribution of genes along a circular genome of length $L$, the probability of observing a distance $d$ less than $xL/2$ between 2 given genes is just $x$. In the null model, the number $M_{ij}(x)$ of genomes with $d \leq x$ thus follows a binomial law $\mathcal{B}(M,x)$.  The probability $\pi_{ij}(x)$ of observing $M_{ij}(x)$ events is therefore $\pi_{ij}(x)=I_x(M_{ij}(x),M'-M_{ij}(x)+1)$, where $I_x(m,n)$ is the regularized incomplete beta function. The least likely and therefore most significant distance $\hat x_{ij}$ between a given pair of genes $ij$, is the one minimizing $\pi_{ij}(x)$, which defines a distance $\hat x_{ij}$ and an associated $p$-value $\hat \pi_{ij}=\pi_{ij}(\hat x_{ij})$.\\
\indent Under the null model, the distribution of $y_{ij}=-\ln \hat \pi_{ij}$ is found to have an exponential tail, $f_0(y)\sim e^{-ay}$, with an exponent $a$ depending on $M'$ (Fig.~\ref{fig:Fig1} and Suppl. Info.). Given a threshold of significance $\pi^*$, we compute the fraction $\s_s$ of significant pairs, with $\hat\pi_{ij}\leq\pi^*$, and estimate the fraction of false positive pairs as $\s_{\rm fp}=\int_{-\ln\pi^*}^\infty f_0(y)\simeq (\pi^*)^a$. Imposing a  false discovery rate $q=\s_{\rm fp}/\s_s$ thus determines the threshold of significance $\pi^*$.\\
\indent To account for the fact genomes may have several chromosomes, be non-circular and have different lengths, we formally circularize linear chromosomes and normalize them to a common length $L$ by setting all distances exceeding $L/2$ to $L/2$. The distance between genes on distinct chromosomes is also set to $L/2$. We take $L = 500 $ kb, but our results are not sensitive to the exact value of this cutoff.\\

\indent To treat pairs of COGs $ij$ with multiple copies (genes), we fix a gene $g_i$ in $i$, count the number $n$ of genes in $j$ at distance less than $xL/2$ and compute the probability of the event as $p(x)=1-(1-x)^n$. The analysis is then performed as for $n=1$ with $\pi_{g_ij}(x)$ now standing for $\pi_{g_ij}(p(x))$, thus defining $\hat\pi_{g_ij}$. We then define $\hat\pi_{ij}$ as the most significant observation when considering successively each gene $g_i$ in $i$, i.e., $\hat\pi_{ij}=\min_{g_i\in i}\{\hat\pi_{g_ij}\}$. Different numbers of genes in $i$ and $j$ may imply $\hat\pi_{ij} \neq \hat\pi_{ji}$. Pairs of proximal COGs are identified by requiring that both $\hat\pi_{ij}\leq\pi^*$ and $\hat\pi_{ji}\leq\pi^*$.\\

\indent {\bf Syntons and pan-syntons} -- For a given gene in a given genome, we build the maximal set of genes (including the gene itself) that are fully interconnected in the pan-network of proximity (clique) and close-by along the genome (Fig.~\ref{fig:Fig2}A); two genes are considered to be close-by if they are separated by less than 50 kb (or less than $\hat x_{ij}$ if $\hat x_{ij} > 50$ kb, which occurs rarely, Fig. S4; taking a smaller value than 50 kb does not affect the results). In cases where several syntons of same size are possible, we take the synton with most significant score, computed by summing the $-\log(\hat \pi_{ij})$ over each pair of genes.\\

\indent {\bf Refinement of COG annotation: cCOGs} --
We iteratively partition the COGs into subsets of "contextually refined COGs" (cCOGs) by spliting them based on the synton to which they belong in each given genome. Each COG $i$ is thus independently reannotated by iterating the following steps:\\		
\indent (1) Identification of COGs with conserved genomic proximity with respect to $i$.\\
\indent (2) Identification of the cliques $c$ to which $i$ belongs in the resulting network of proximal pairs of COGs. The cliques correspond a priori to incompatible genomic contexts, i.e., contexts found in different genomes.\\
\indent (3) The maximal cliques $c$ to which $i$ belongs are scored independently in each genome $g$ by $\s_{ic}^{(g)}=\sum_j-\log (\hat \pi_{ij})$, where the sum is over the COGs $j$ that are both in clique $c$ and in the context of $i$ in $g$. In every genome, the clique with the best score is associated to $i$, which is hence partitioned in cCOGs (see Fig.~S12A).\\
\indent (4) A consensus genomic context is computed for each $i$ annotated in the previous step. COGs not annotated in the previous step are then annotated using the best-matching context of these COGs.\\
\indent We stop the iterations when 99\% of the COGs have their annotation unchanged in two successive iterations. For simplicity, the identification of the context of a COG, or of a cCOG, is always done using the original COGs. Note also that because the identification of maximal cliques is computationally demanding for large pan-networks, we identify the maximal cliques of the subnetwork associated to every COG, rather than the maximal cliques in the full network (see Fig.~S12A).\\
\indent At the most generic phylogenetic depth ($\delta=0.91, M'=10$), we take as input the original COGs. For smaller values of $\delta$, we start from the cCOGs obtained at a value of $\delta$ slightly larger. Eventually, we thus obtain many more cCOGs than COGs and, hence, larger pan-networks of proximity: $\sim$5200 cCOGs with 1300 proximal pairs at $\delta=0.91$ ($M'=10$), $\sim 48000$ cCOGs with 131000 pairs at $\delta = 0.68$ ($M'=304$). In any case, the size distribution of (pan-)syntons always leads to two types of syntons, with an exponential distribution for type A syntons (Fig.~S12B).

\section{Acknowledgments}
We thank Toni Gabald\'on, Bahram Houchmandzadeh and Kim Reynolds for helpful comments. I.J. thanks Fran\c{c}ois K\'ep\`es for his input at an early stage of this work. This work was supported by a Novartis grant (to I.J.) and by ANR grant 'CoEvolInterProt' (to O.R.).

\bibliographystyle{unsrt}

\clearpage
\setcounter{figure}{0}
\onecolumngrid
\appendix
\newpage

\section{SUPPLEMENTARY INFORMATION}

\subsection{Measure of context divergence}

The context divergence $\delta$ between any two genomes is computed as $\delta = 1-f$, where $f$ is the average fraction of common COGs in the context of a selection of 10 genes. The context of a gene is defined as the COGs located within 20 kb of this gene, and the 10 selected genes are associated to the COGs 126G, 173J, 202K, 2255L, 481M, 497L, 541U, 544O, 556L, 1158K. These COGs are taken from a list of genes shown to report phylogenetic distances between bacterial strains (Table~2 in \cite{Zeigler:2003hg}), with the additional constraint that they comprise a single copy in most of the 1108 genomes of our dataset.

Out of these 10 genes, only 202K shows a particularly conserved context. Comparing the context divergence obtained from two distinct sets of 5 genes, with 202K in common only, shows a good linear relation between the two estimations (Fig.~S\ref{fig:self}). This self-averaging property indicates that using these 10 genes is suitable for measuring the context divergence between pairs of genomes.

\subsection{Measure of sequence divergence}

The same 10 genes are used to compute a measure of sequence divergence $\delta_s$ between any two genomes as $\delta_s = 1-f_s$, where $f_s$ is the average fraction of common amino acids between the 10 genes, after alignment of their amino acid sequences. $\delta_s$ and $\delta$ are related as indicated in Fig.~S\ref{fig:delta}.

\subsection{Distribution of $p$-values for the null model of proximity tendencies}

A null model is defined by assuming that the positions of genes are drawn from an uniform distribution, independently in each of $M$ genomes. For each pair $ij$ of genes, the number of genomes for which the distance between $i$ and $j$ is smaller than $x$ is translated into a $p$-value $\pi_{ij}(x)$ (Methods). As any $p$-value, $\pi_{ij}(x)$ is uniformly distributed over the pairs $ij$ for each given $x$, or, equivalently, $y_{ij}=-\log \pi_{ij}(x)$ is exponentially distributed, $\psi(y)=e^{-y}$.

The quantity $\hat\pi_{ij}=\min_x\pi_{ij}(x)$ is not a $p$-value, but numerical simulations show that the tail of the distribution of $\hat y_{ij}=-\log \hat\pi_{ij}(x)$ is exponentially distributed, $\psi(\hat y)\sim e^{-a \hat y}$, with an exponent $a$ that depends on $M$ (Fig.~S\ref{fig:null}). Taking for $M$ the effective number of genomes $M'$ gives this exponent as a function of the context divergence $\delta$.

\subsection{Significance of the exponential distribution of type A synton sizes}

\subsubsection{Randomization of gene positions}

To support the non-trivial nature of the exponential distribution of type A synton sizes, we repeated the analysis after randomly permuting the labels of the COGs of the pan-syntons of certain sizes (obtained at $\delta = 0.83$). Fig.~S\ref{fig:rando} shows the results, where the blue points correspond to randomizing the pan-syntons of size 3-6 and the red points those of size 4-6. In any case, the exponential nature of the distribution is lost.

\subsubsection{Varying false discovery rates}

Using a more stringent false discovery rate than $q=0.05$ reduces the statistics but does not affect the conclusion that pan-syntons can be divided into two types according to their size, with the size of the smaller ones being exponentially distributed (Fig.~S\ref{fig:varfdr}).

\begin{table}[b]
\renewcommand{\tablename}{Table S}
\begin{center}
	\begin{tabular}{ | p{0.45\linewidth} |p{0.15\linewidth}|p{0.35\linewidth}|}
	\hline	
	Connected component & Biological function & Example of syntons \\
	\hline
 	1005C 1007C 1008C 1009CP 1034C 1143C 1894C 1905C 377C 649C 713C 838C 839C 852C
& NADH:ubiquinone oxidoreductase
	&1005C 1007C 1008C 1009CP 1034C 1143C 1894C 1905C 377C 649C 713C 838C 839C 852C\\
	\hline
	1181M 1589M 2001S 206D 275M 3116D 472M 707M 768M 769M 770M 771M 772D 773M 812M 849D 
	& Cell division / Cell envelope biogenesis
	&1181M 1589M 2001S 206D 275M 472M 707M 768M 769M 770M 771M 772D 773M 812M 849D\\
	\cline{3-3}
	&&1181M 1589M 2001S 206D 275M 3116D 472M 707M 768M 769M 770M 771M 772D 773M 849D \\
	\hline
	1157NU 1191K 1291N 1298NU 1317NU 1338NU 1360N 1377NU 1419N 1536N 1558N 1580N 1582N 1677NU 1684NU 1749N 1766NU 1776NT 1815N 1843N 1868N 1886NU 1987NU 2063N 3144N 3190N 455D 4786N
	& Flagellum
	&1291N 1298NU 1360N 1338NU 1377NU 1419N 1580N 1582N 1684NU 1749N 1843N 1868N 1886NU 1987NU 3144N 3190N 455D\\
	\cline{3-3}
	&&1298NU 1338NU 1377NU 1536N 1558N 1677NU 1684NU 1749N 1766NU 1843N 1868N 1886NU 1987NU 4786N\\
	\hline
	100J 101J 102J 103J 1841J 185J 186J 197J 198J 199J 200J 201U 202K 203J 222J 244J 24J 250K 255J 256J 257J 361J 480J 48J 49J 50J 51J 522J 563F 690U 80J 81J 85K 86K 87J 88J 89J 90J 91J 92J 93J 94J 96J 97J 98J 99J& Ribosome / RNA polymerase&100J 101J 102J 103J 200J 201U 202K 203J 255J 256J 361J 563F 94J 96J 97J 98J 99J\\
	\cline{3-3}
	&&100J 1841J 185J 186J 197J 198J 199J 200J 201U 202K 203J 255J 256J 480J 48J 49J 50J 51J 522J 85K 86K 87J 88J 89J 90J 91J 92J 93J 94J 96J 97J 98J 99J\\
	\hline
	\end{tabular}	
\caption{COG composition of type B pan-syntons, defined at $\delta = 0.83$ as pan-syntons containing more than 13 COGs. The left column indicates the composition of the four connected components that are obtained when restraining the pan-network of proximity to the set of COGs that are included in type B syntons (graphs below); the biological function associated  with each component is indicated in the center. The right column provides two examples of syntons for each component (only one in the case of NADH:ubiquinone oxidoreductase). The same biological functions are found when the syntons are determined using cCOGs. In this case, we also find the F0F1-type ATP synthase machinery; together with the ubiquinone oxidoreductase, it corresponds to the core of the ATP-producing respiratory macro-molecular complex.\label{tab:syntonB}}
\end{center}
\end{table}

\begin{figure}
\centerline{\includegraphics[width=0.6\linewidth]{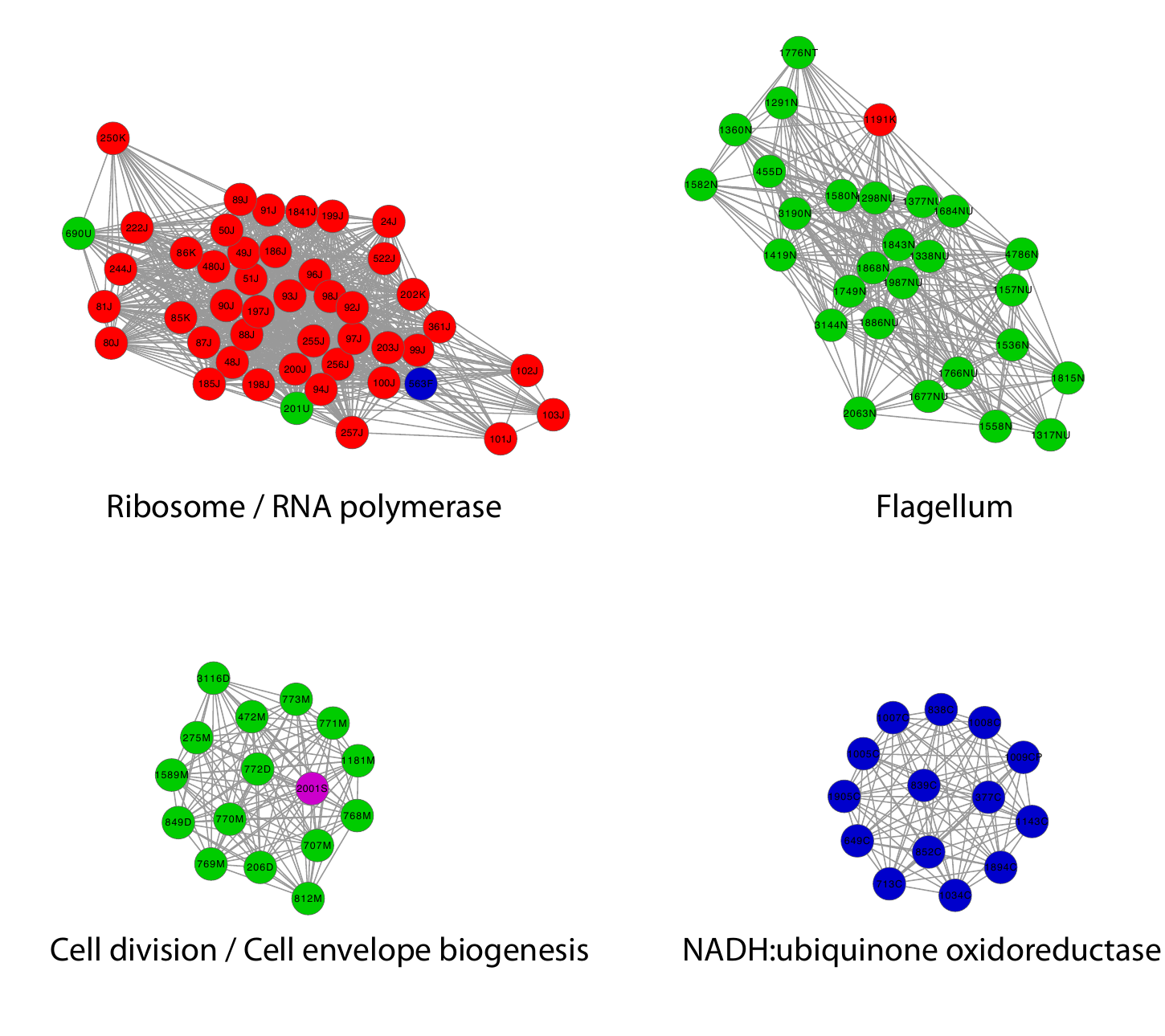}}
\end{figure}

\clearpage

\begin{figure}
\renewcommand{\figurename}{FIG. S}
\centering
\includegraphics[width=.5\linewidth]{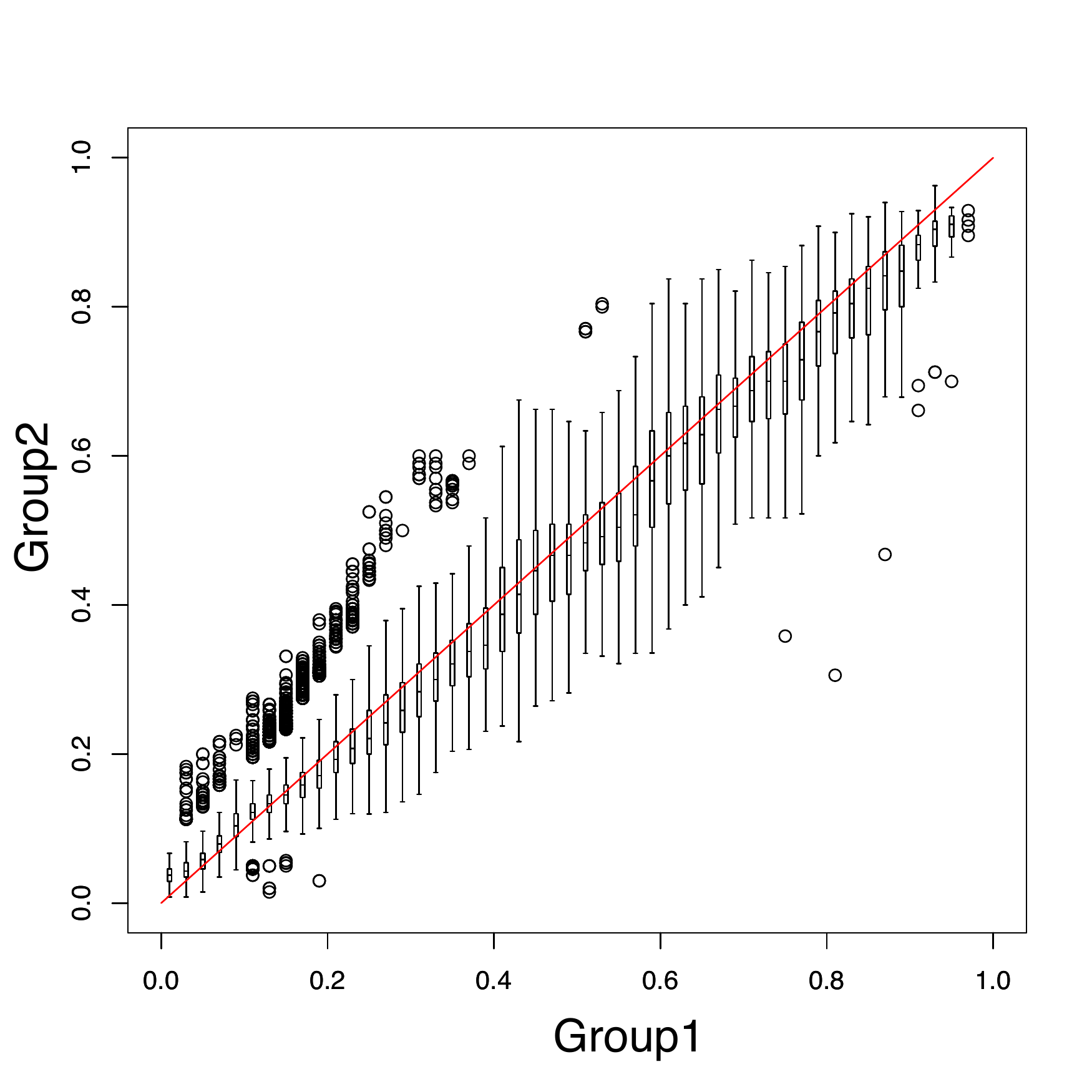}
\caption{Comparison of context similarity $f=1-\delta$ computed from two different groups of 5 genes (box plot). All genome pairs formed from the 1108 genomes of our dataset are reported. \label{fig:self}}
\end{figure}

\begin{figure}
\renewcommand{\figurename}{FIG. S}
\centering
\includegraphics[width=0.5\linewidth]{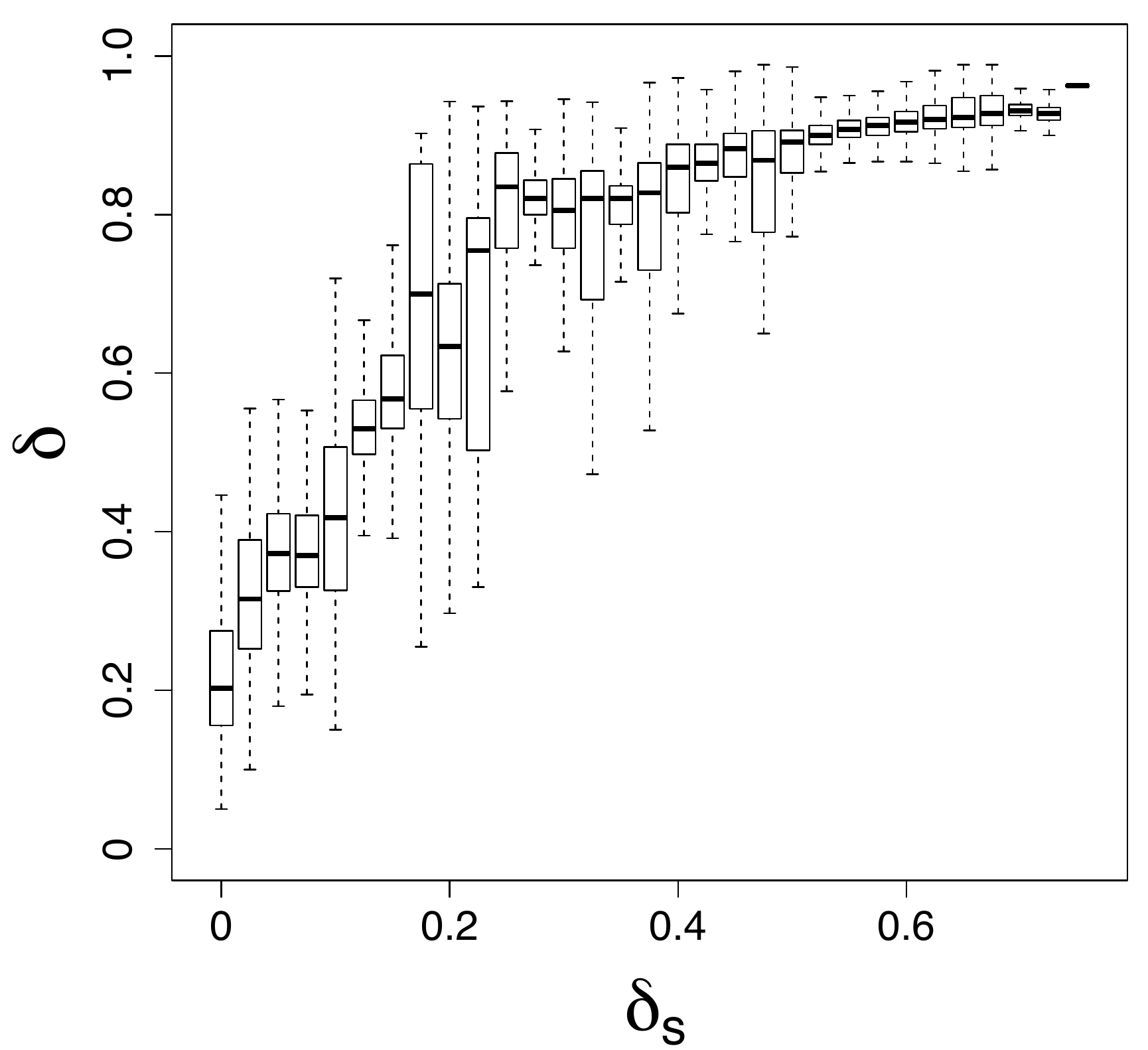}
\caption{Relation between the inter-genomic distance $\delta$ based on context similarity and the inter-genomic distance $\delta_s$ based on sequence similarity (box plot). The two are correlated, but context similarity vanishes before sequence similarity and $\delta_s$ therefore describes large divergences  better. \label{fig:delta}}
\end{figure}

\clearpage

\begin{figure}
\renewcommand{\figurename}{FIG. S}
\centerline{\includegraphics[width=0.7\linewidth]{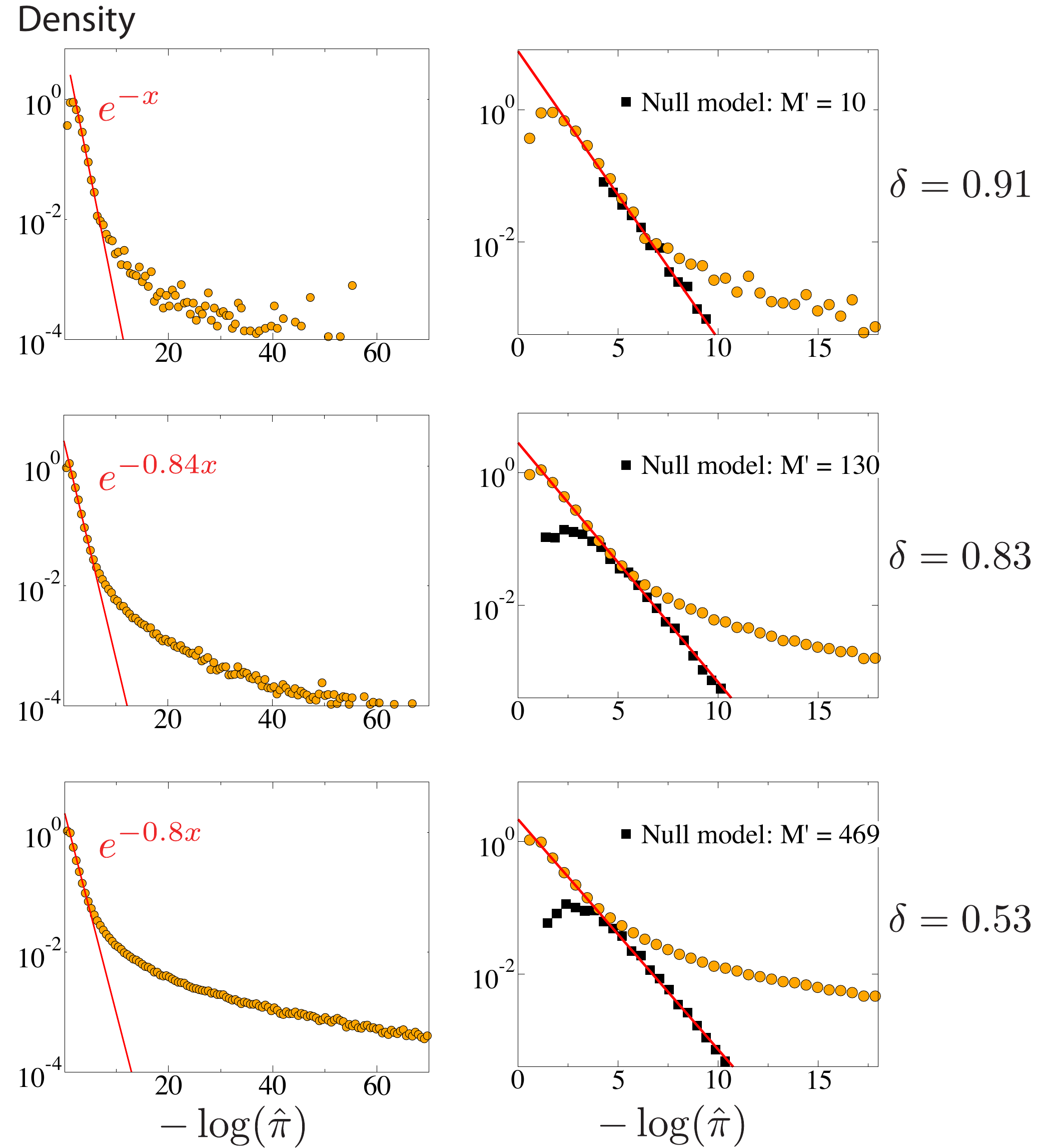}}
\caption{Probability density of $-\log(\hat \pi)$ for the empirical data (orange circles) at three phylogenetic depths: $\delta=0.91,0.83,0.53$. Left panels: For small enough values of $-\log(\hat \pi)$, the density decays exponentially with $-\log(\hat \pi)$ (red lines). The deviation from an exponential at large values indicates the conservation of co-localization. For the null model where gene positions are randomized (black squares, right panels), with as number of genomes the effective number $M'$ corresponding to $\delta$ ($M'=10,130,469$, respectively), the exponential decay extends to larger values of $-\log(\hat \pi)$.
\label{fig:null}}
\end{figure}

\begin{figure}
\renewcommand{\figurename}{FIG. S}
\centerline{\includegraphics[width=0.39\linewidth]{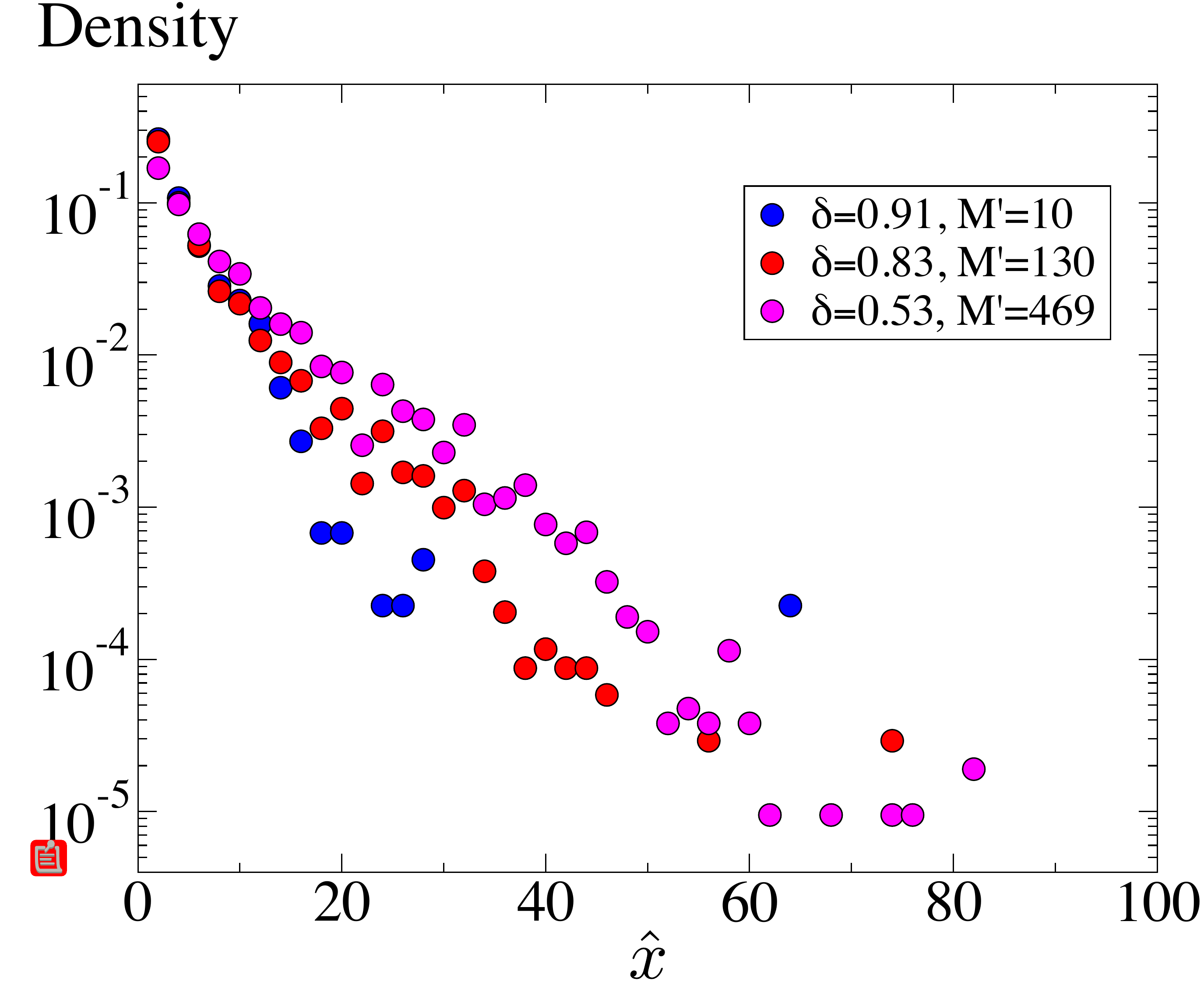}}
\caption{Distribution of the characteristic distances $\hat x_{ij}$ (in kb) associated with $\hat \pi$ (see Fig.~1B in main text).
\label{fig:distances}}
\end{figure}

\clearpage

\begin{figure}
\renewcommand{\figurename}{FIG. S}
\centering
\includegraphics[width=0.75\linewidth]{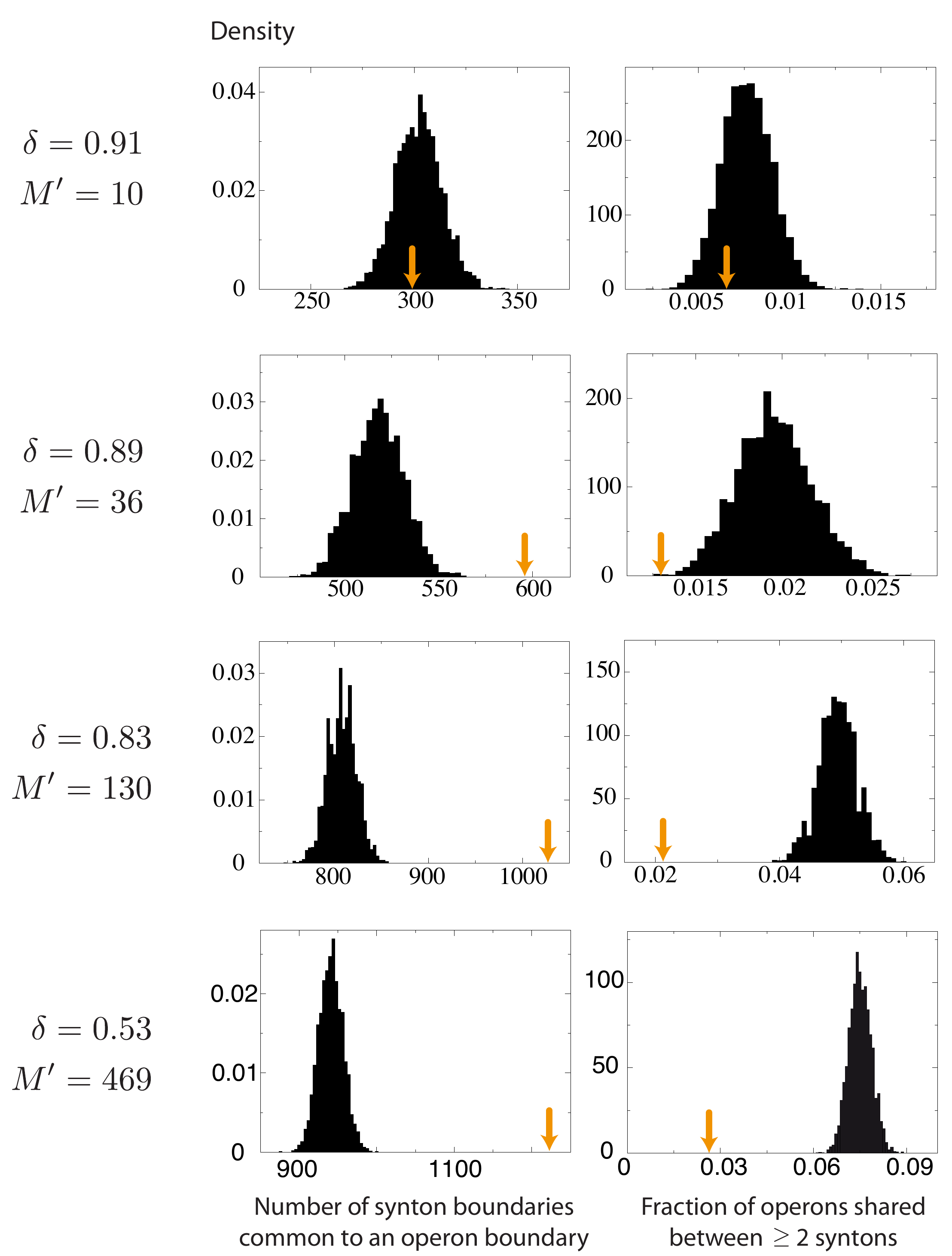}
\caption{Significance of the observed number of synton boundaries common to an operon boundary and of the fraction of operons shared between more than 2 syntons when comparing with randomly rotated genomes. This figure extends Fig.~2C to different values of the phylogenetic depth $\delta$.}\label{fig:}
\end{figure}

\clearpage

\begin{figure}[t]
\renewcommand{\figurename}{FIG. S}
\centering
\includegraphics[width=.4\linewidth]{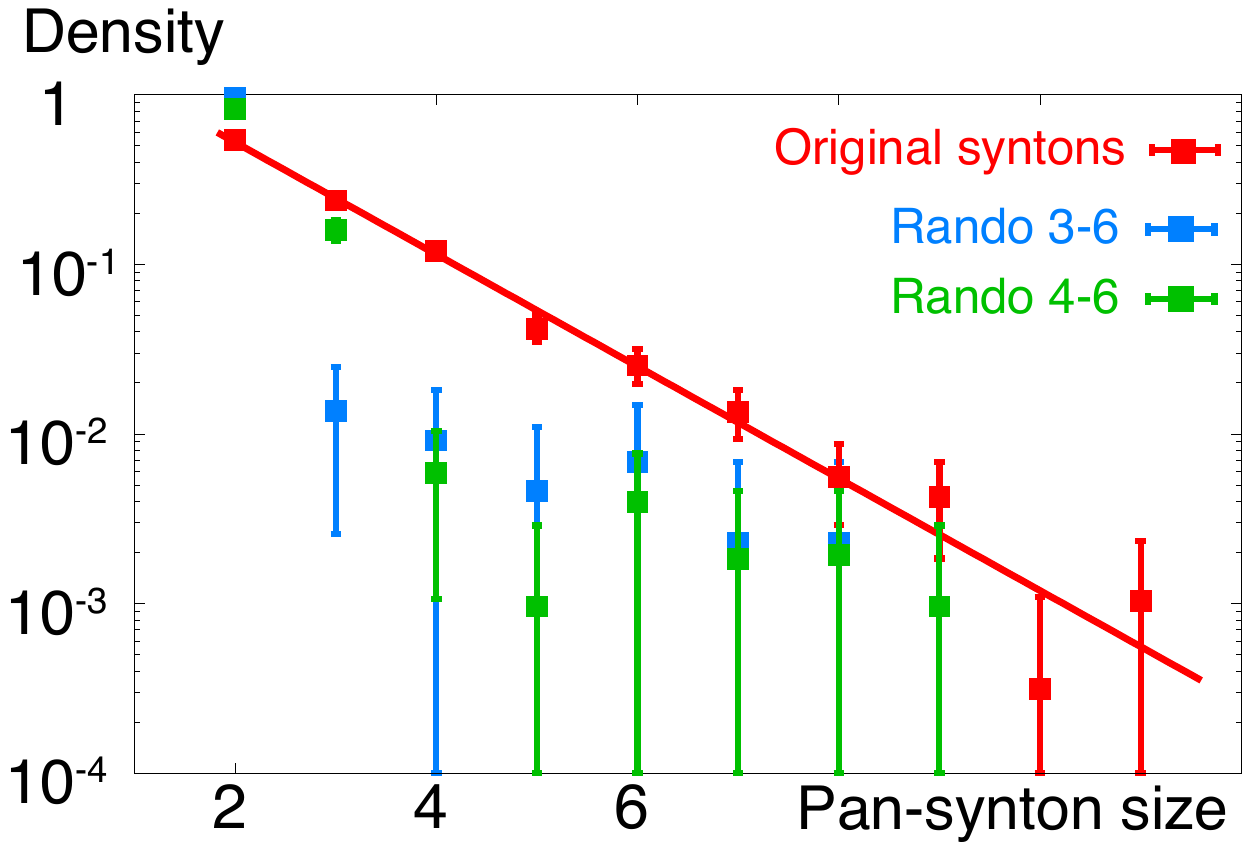}
\caption{Effect of randomization on the distribution of pan-synton sizes ($\delta = 0,83; M'=130$). Red points show the original distribution of type A pan-syntons  (Fig.~4 in main text). The blue and green curves show the distributions obtained after reshuffling the positions of the COGs belonging to pan-synton of sizes 3 to 6 and 4 to 6, respectively. Reshuffling all COG positions leads to a very small graph that contains only cliques of size 2 (not shown).\label{fig:rando}}
\end{figure}

\begin{figure}[t]
\renewcommand{\figurename}{FIG. S}
\centerline{\includegraphics[width=.4\linewidth]{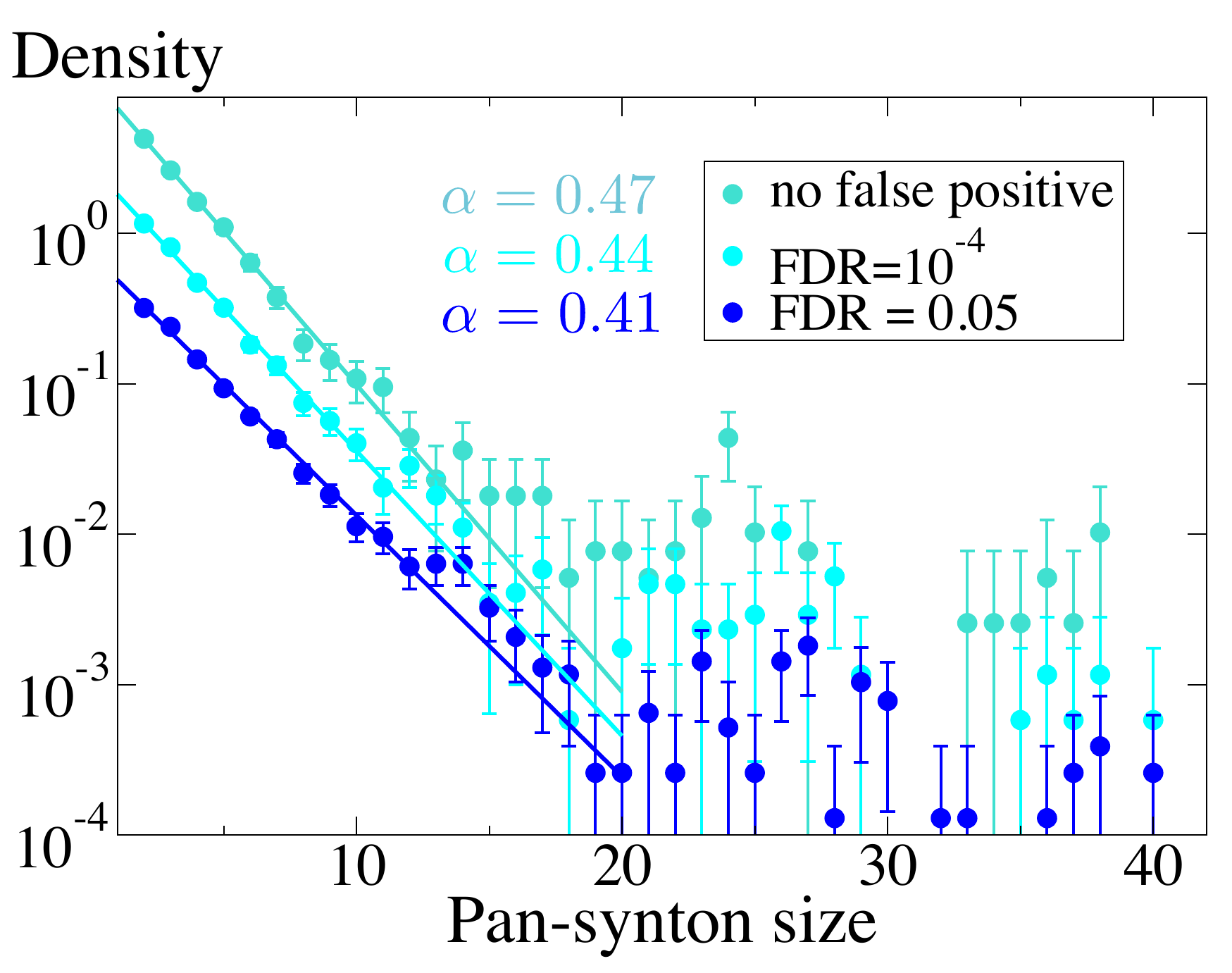}}
\caption{Distribution of pan-synton sizes at $\delta = 0.53$ ($M'=469$) for different false discovery rates (FDRs), showing again the presence of two types of pan-syntons with an exponential decay for type A pan-syntons. The value of the exponent $\alpha$ for the exponential decay is indicated. FDRs equal to $0.05$ and $10^{-4}$ corresponds respectively to $\pi^*=10^{-4}$ and $\pi^* = 10^{-6}$; an expected absence of false positives correspond to  $\pi^* = 10^{-9}$. For clarity, curves have been shifted along the $y$-axis for $\text{FDR} = 10^{-4}$ and for the absence of false positives.\label{fig:varfdr}}
\end{figure}

\begin{figure}[t]
\renewcommand{\figurename}{FIG. S}
\centerline{\includegraphics[width=.4\linewidth]{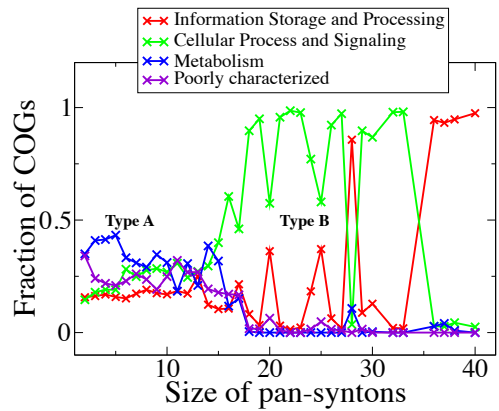}}
\caption{Compositions of pan-syntons in terms of the four main functional classes of COGs \cite{Tatusov:2000tu}, as a function of their size. For type B pan-syntons, the categories "information storage and processing" (translation/transcription  machinery) and "cellular process and signaling" (flagellum, cell division machineries) are over-represented. For type A pan-syntons, all functional classes are equally represented. \label{fig:funcrep}}
\end{figure}

\clearpage

\begin{figure}
\renewcommand{\figurename}{FIG. S}
\centerline{\includegraphics[width=0.4\linewidth]{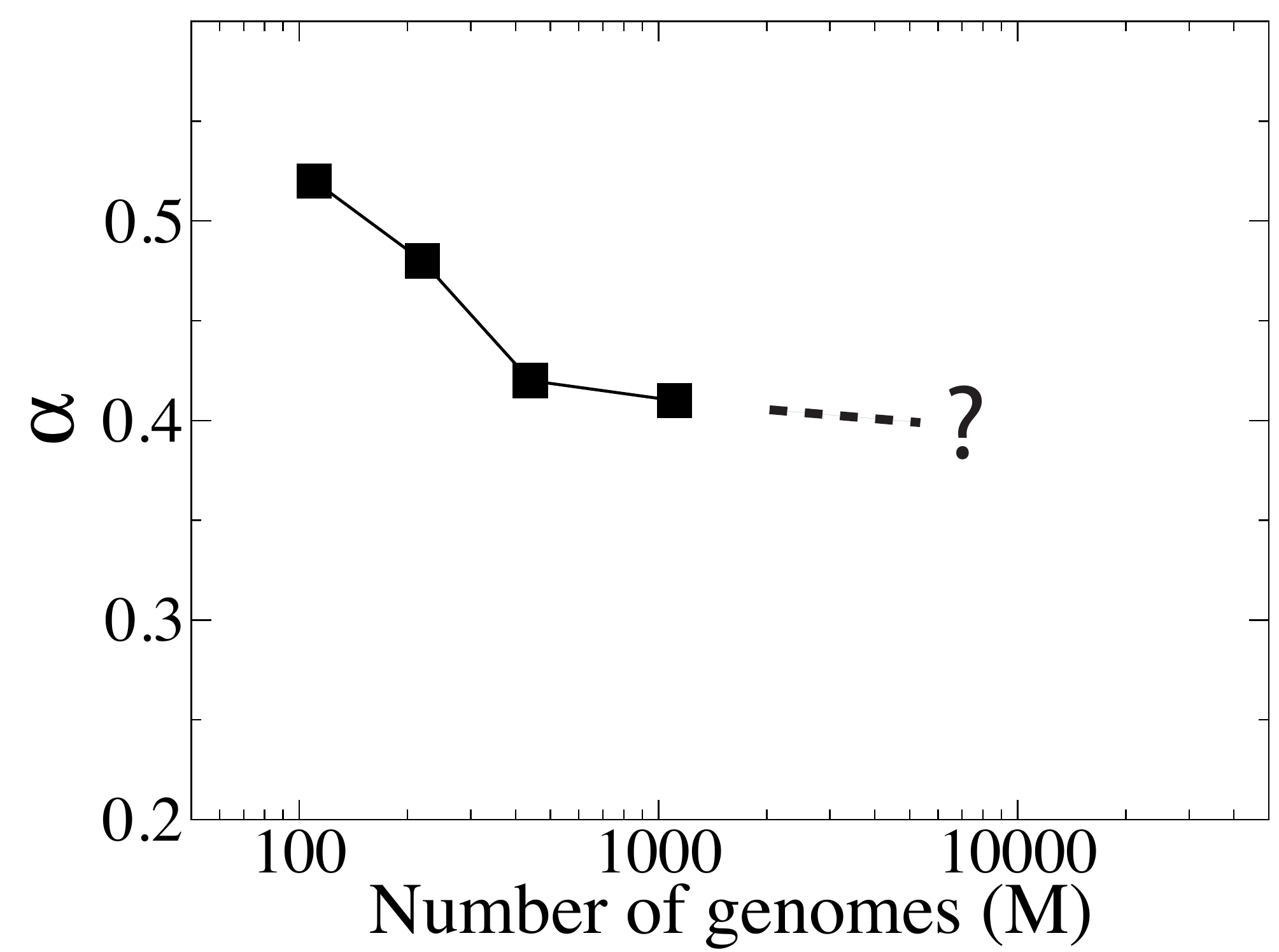}}
\caption{Exponent $\alpha$ for the exponential distribution of the size of type A pan-syntons at $\delta = 0.53$ ($\text{FDR} = 0.05$), for different numbers of genomes in the dataset; datasets with $M<1108$ genomes were obtained by resampling the original dataset of $M=1108$ strains (rightmost point). This analysis is consistent with a saturation of the value of $\alpha$ with the number of genomes, which future analyses including more genomes will be able to confirm or infirm.
\label{fig:distances}}
\end{figure}

\begin{figure}
\renewcommand{\figurename}{FIG. S}
\centerline{\includegraphics[width=0.7\linewidth]{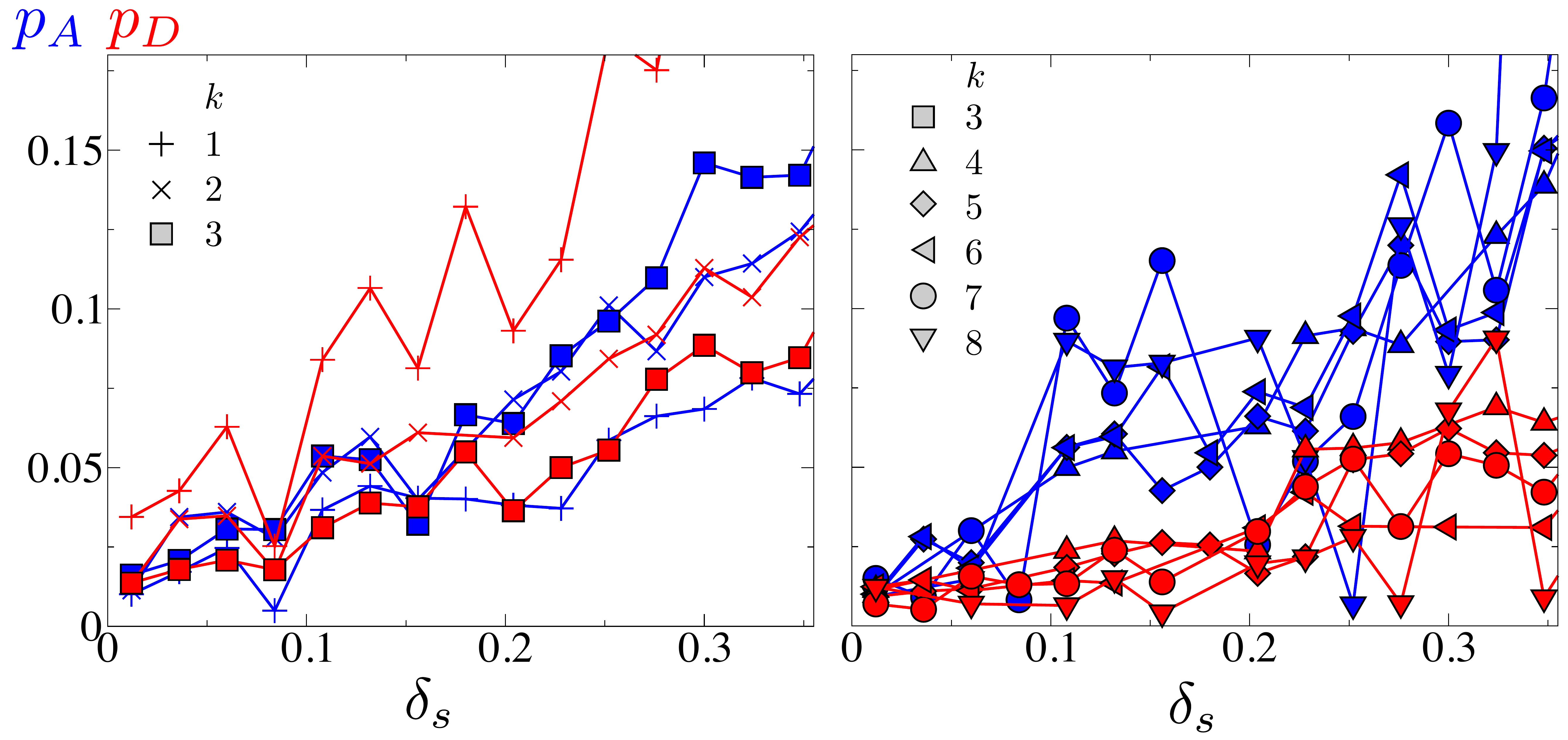}}
\caption{\label{fig:pApD} Generalization of Fig. 6C to other values of $k$.}
\end{figure}

\begin{figure}
\renewcommand{\figurename}{FIG. S}
\centerline{\includegraphics[width=0.35\linewidth]{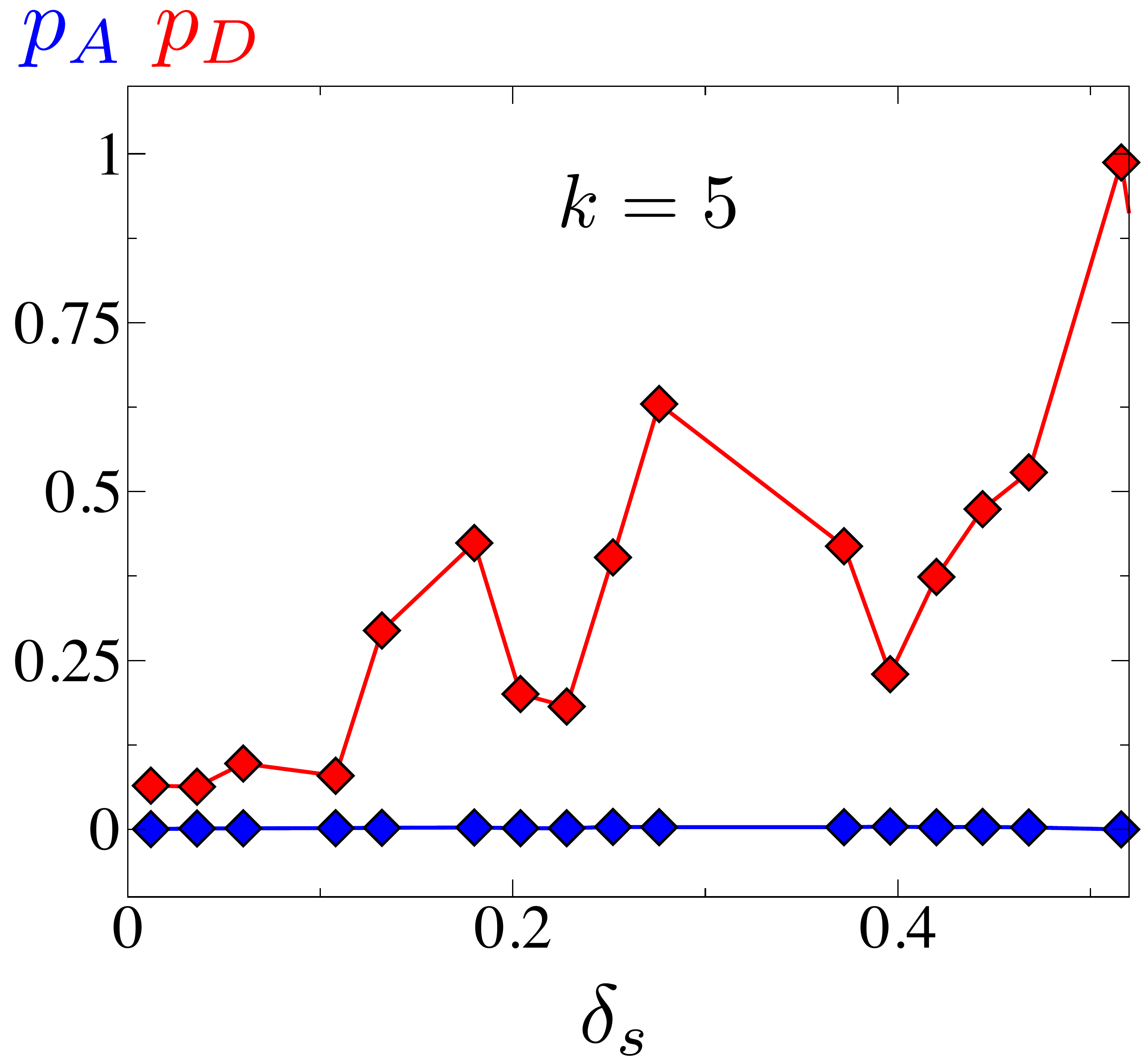}}
\caption{\label{fig:pApDneutral} Probabilities of aggregation ($p_A$) and disaggregation ($p_D$) obtained from genomic contexts that are not conserved. The graph displays the results for the context of synteny units that contain 5 genes ($k=5$ in Fig.~6 of main text); for conserved contexts, the dynamics  is dominated by aggregation events (see Fig.~6 of main text). Here, we see that $p_A$ is negligible for all values of $\delta_s$; the same observation holds for all sizes of the synteny units (points not represented correspond to values of $p_A$ that were found to be slightly negative).}
\end{figure}

\begin{figure}[t]
\vspace*{.05in}
\renewcommand{\figurename}{FIG. S}
\centerline{\includegraphics[width=0.95\linewidth]{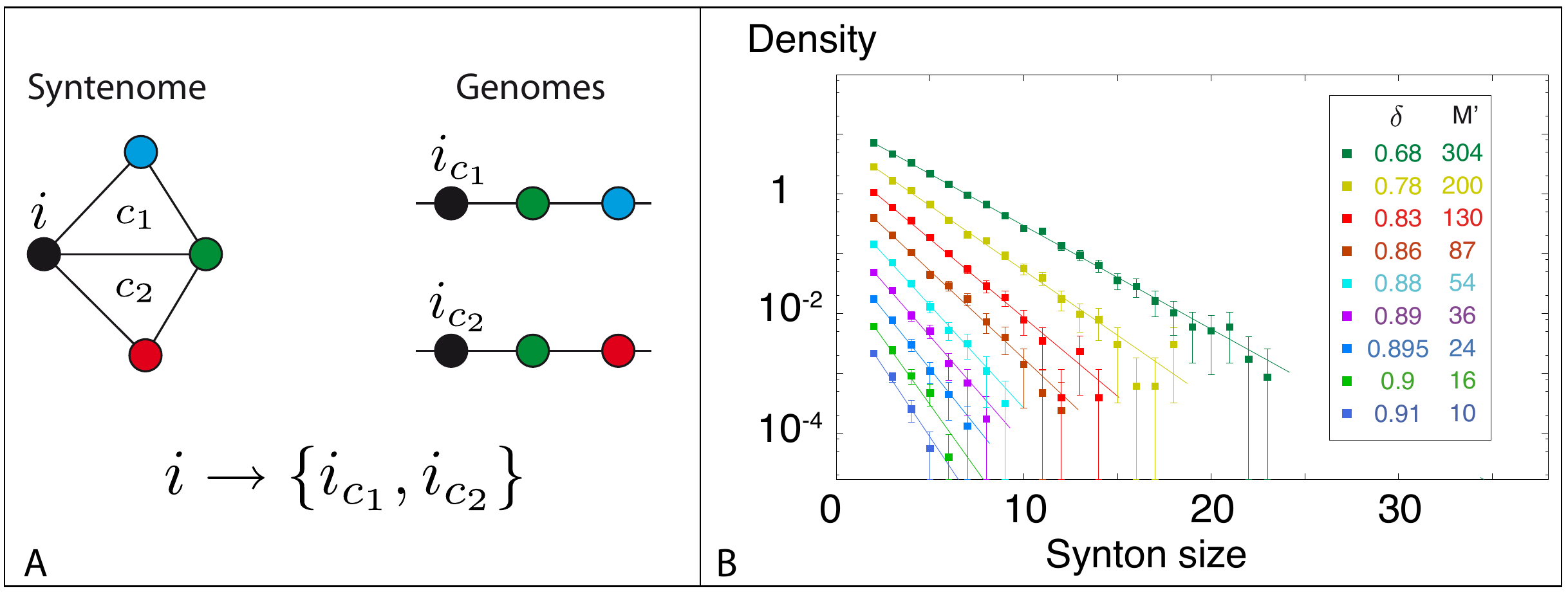}}	
\caption{{\bf(A)} A COG $i$ is partitioned in two cCOGs ($i_{c_1}$ and $i_{c_2}$) due to the presence of two maximal cliques ($c_1$ and $c_2$) in the subnetwork associated with $i$ (left), corresponding to genome organizations found in different genomes (right). {\bf(B)} Size distributions of type A pan-syntons obtained from the pan-network of proximity associated with the cCOGs.\label{fig:syntons_typeA}}	
\end{figure}

\end{document}